\documentclass[ALICE,manyauthors]{cernphprep}

\usepackage[comma,square,numbers,sort&compress]{natbib}
\usepackage{hyperref}
\usepackage{lineno}
\usepackage{xspace}
\usepackage{xcolor}
\usepackage[yyyymmdd]{datetime}
\usepackage[T1]{fontenc}

\newcommand{\y}{\ensuremath{y}\xspace}
\newcommand{\jpsi}{\ensuremath{{\rm J}/\psi}\xspace}
\newcommand{\meanptjpsi}{\ensuremath{\langle p^{\rm{J}/\psi}_{\rm{T}}\rangle}\xspace}
\newcommand{\dnchdeta}{\ensuremath{{\rm d}N_{\rm{ch}}/{\rm d}\eta}\xspace}

\newcommand{\avgdnchdeta}{\ensuremath{{{\rm d}N_{\rm ch}/{\rm d}\eta}/{\langle {\rm d}N_{\rm ch}/{\rm d}\eta \rangle}}\xspace}
\newcommand{\Nch}{\ensuremath{N_{\rm ch}}\xspace}
\newcommand{\zSPDv}{\ensuremath{z^{\rm SPD}_{\rm vtx}}\xspace}
\newcommand{\Ntrk}{\ensuremath{N_{\rm trk}}\xspace}
\newcommand{\Ntrkcorr}{\ensuremath{N_{\rm trk}^{\rm corr}}\xspace}
\newcommand{\meanNtrkcorr}{\ensuremath{\langle N_{\rm trk}^{\rm corr}\rangle}\xspace}
\newcommand{\acceff}{\ensuremath{A\times\epsilon}\xspace}
\newcommand{\ptmu}{\ensuremath{\langle p^{\rm{\mu^{+}\mu^{-}}}_{\rm T}\rangle}\xspace}
\begin{document}
%

\newcommand{\pp}           {pp\xspace}
\newcommand{\ppbar}        {\mbox{$\mathrm {p\overline{p}}$}\xspace}
\newcommand{\XeXe}         {\mbox{Xe--Xe}\xspace}
\newcommand{\PbPb}         {\mbox{Pb--Pb}\xspace}
\newcommand{\pA}           {\mbox{pA}\xspace}
\newcommand{\pPb}          {\mbox{p--Pb}\xspace}
\newcommand{\AuAu}         {\mbox{Au--Au}\xspace}
\newcommand{\dAu}          {\mbox{d--Au}\xspace}

\newcommand{\s}            {\ensuremath{\sqrt{s}}\xspace}
\newcommand{\snn}          {\ensuremath{\sqrt{s_{\mathrm{NN}}}}\xspace}
\newcommand{\pt}           {\ensuremath{p_{\rm T}}\xspace}
\newcommand{\meanpt}       {$\langle p_{\mathrm{T}}\rangle$\xspace}
\newcommand{\ycms}         {\ensuremath{y_{\rm CMS}}\xspace}
\newcommand{\ylab}         {\ensuremath{y_{\rm lab}}\xspace}
\newcommand{\etarange}[1]  {\mbox{$\left | \eta \right |~<~#1$}}
\newcommand{\yrange}[1]    {\mbox{$\left | y \right |~<~#1$}}
\newcommand{\dndy}         {\ensuremath{\mathrm{d}N_\mathrm{ch}/\mathrm{d}y}\xspace}
\newcommand{\dndeta}       {\ensuremath{\mathrm{d}N_\mathrm{ch}/\mathrm{d}\eta}\xspace}
\newcommand{\avdndeta}     {\ensuremath{\langle\dndeta\rangle}\xspace}
\newcommand{\dNdy}         {\ensuremath{\mathrm{d}N_\mathrm{ch}/\mathrm{d}y}\xspace}
\newcommand{\Npart}        {\ensuremath{N_\mathrm{part}}\xspace}
\newcommand{\Ncoll}        {\ensuremath{N_\mathrm{coll}}\xspace}
\newcommand{\dEdx}         {\ensuremath{\textrm{d}E/\textrm{d}x}\xspace}
\newcommand{\RpPb}         {\ensuremath{R_{\rm pPb}}\xspace}

\newcommand{\nineH}        {$\sqrt{s}~=~0.9$~Te\kern-.1emV\xspace}
\newcommand{\seven}        {$\sqrt{s}~=~7$~Te\kern-.1emV\xspace}
\newcommand{\twoH}         {$\sqrt{s}~=~0.2$~Te\kern-.1emV\xspace}
\newcommand{\twosevensix}  {$\sqrt{s}~=~2.76$~Te\kern-.1emV\xspace}
\newcommand{\five}         {$\sqrt{s}~=~5.02$~Te\kern-.1emV\xspace}
\newcommand{\twosevensixnn}{$\sqrt{s_{\mathrm{NN}}}~=~2.76$~Te\kern-.1emV\xspace}
\newcommand{\fivenn}       {$\sqrt{s_{\mathrm{NN}}}~=~5.02$~Te\kern-.1emV\xspace}
\newcommand{\LT}           {L{\'e}vy-Tsallis\xspace}
\newcommand{\GeVc}         {Ge\kern-.1emV/$c$\xspace}
\newcommand{\MeVc}         {Me\kern-.1emV/$c$\xspace}
\newcommand{\TeV}          {Te\kern-.1emV\xspace}
\newcommand{\GeV}          {Ge\kern-.1emV\xspace}
\newcommand{\MeV}          {Me\kern-.1emV\xspace}
\newcommand{\GeVmass}      {Ge\kern-.2emV/$c^2$\xspace}
\newcommand{\MeVmass}      {Me\kern-.2emV/$c^2$\xspace}
\newcommand{\lumi}         {\ensuremath{\mathcal{L}}\xspace}

\newcommand{\ITS}          {\rm{ITS}\xspace}
\newcommand{\TOF}          {\rm{TOF}\xspace}
\newcommand{\ZDC}          {\rm{ZDC}\xspace}
\newcommand{\ZDCs}         {\rm{ZDCs}\xspace}
\newcommand{\ZNA}          {\rm{ZNA}\xspace}
\newcommand{\ZNC}          {\rm{ZNC}\xspace}
\newcommand{\SPD}          {\rm{SPD}\xspace}
\newcommand{\SDD}          {\rm{SDD}\xspace}
\newcommand{\SSD}          {\rm{SSD}\xspace}
\newcommand{\TPC}          {\rm{TPC}\xspace}
\newcommand{\TRD}          {\rm{TRD}\xspace}
\newcommand{\VZERO}        {\rm{V0}\xspace}
\newcommand{\VZEROA}       {\rm{V0A}\xspace}
\newcommand{\VZEROC}       {\rm{V0C}\xspace}
\newcommand{\Vdecay} 	   {\ensuremath{V^{0}}\xspace}

\newcommand{\ee}           {\ensuremath{e^{+}e^{-}}} 
\newcommand{\pip}          {\ensuremath{\pi^{+}}\xspace}
\newcommand{\pim}          {\ensuremath{\pi^{-}}\xspace}
\newcommand{\kap}          {\ensuremath{\rm{K}^{+}}\xspace}
\newcommand{\kam}          {\ensuremath{\rm{K}^{-}}\xspace}
\newcommand{\pbar}         {\ensuremath{\rm\overline{p}}\xspace}
\newcommand{\kzero}        {\ensuremath{{\rm K}^{0}_{\rm{S}}}\xspace}
\newcommand{\lmb}          {\ensuremath{\Lambda}\xspace}
\newcommand{\almb}         {\ensuremath{\overline{\Lambda}}\xspace}
\newcommand{\Om}           {\ensuremath{\Omega^-}\xspace}
\newcommand{\Mo}           {\ensuremath{\overline{\Omega}^+}\xspace}
\newcommand{\X}            {\ensuremath{\Xi^-}\xspace}
\newcommand{\Ix}           {\ensuremath{\overline{\Xi}^+}\xspace}
\newcommand{\Xis}          {\ensuremath{\Xi^{\pm}}\xspace}
\newcommand{\Oms}          {\ensuremath{\Omega^{\pm}}\xspace}
\newcommand{\degree}       {\ensuremath{^{\rm o}}\xspace}

\begin{titlepage}
\PHyear{2021}       
\PHnumber{268}      
\PHdate{17 December}  

\title{Forward rapidity J/$\psi$ production as a function of charged-particle multiplicity in \pp collisions at $\s =$ 5.02
and 13 TeV}
\ShortTitle{Forward rapidity J/$\psi$ production as a function of charged-particle multiplicity}   

\Collaboration{ALICE Collaboration\thanks{See Appendix~\ref{app:collab} for the list of collaboration members}}
\ShortAuthor{ALICE Collaboration} 

\begin{abstract}
  
The production of \jpsi is measured as a function of charged-particle multiplicity at forward rapidity in proton--proton (pp) collisions at center-of-mass energies $\s =$ 5.02 and 13 TeV. 
The \jpsi mesons are reconstructed via their decay into dimuons in the rapidity interval (2.5 $< y <$ 4.0), whereas the charged-particle multiplicity density (\dnchdeta) is measured at midrapidity $(|\eta| < 1)$. The production rate as a function of multiplicity is reported as the ratio of the yield in a given multiplicity interval to the multiplicity-integrated one. This observable shows a linear increase with charged-particle multiplicity normalized to the corresponding average value for inelastic events (\avgdnchdeta), at both the colliding energies. 
Measurements are compared with available ALICE results at midrapidity and theoretical model calculations.
First measurement of the mean transverse momentum (\meanpt) of \jpsi in pp collisions exhibits an increasing trend as a function of \avgdnchdeta showing a saturation towards high charged-particle multiplicities. 

\end{abstract}
\end{titlepage}

\setcounter{page}{2} 


\section{Introduction}

The study of charmonia, bound states of a charm and anti-charm quark (${\rm c}\overline{\rm c}$), production in hadronic collisions represents a stringent test for theory of the strong interaction, quantum chromodynamics (QCD). 
While the production of heavy quark-antiquark pair can be calculated within pQCD, their evolution into a bound colorless ${\rm q}\overline{\rm q}$ pair is a non-perturbative process. 
Different theoretical approaches exist, which mainly differ in the treatment of the (non-perturbative) bound state formation~\cite{Fritzsch:1977ay,Amundson:1996qr,Baier:1981uk,Bodwin:1994jh}. 
Inclusive differential studies of $\jpsi$ production as a function of transverse momentum ($\pt$) and rapidity ($y$) have been performed at the LHC in pp collisions at center-of-mass energies ranging from $\sqrt{s}=2.76$~TeV up to 13 TeV~\cite{Aamodt:2011gj,Aaij:2015rla,Abelev:2012kr,Abelev:2014qha,Adam:2015rta,Acharya:2017hjh}. 
The measurements are described by the sum of non-relativistic quantum chromodynamics (NRQCD) calculations for the prompt component, and fixed-order next-to-leading logarithm calculations for the contribution arising from beauty-hadron decays (referred as ''non-prompt'')~\cite{Acharya:2017hjh}. 
Charmonium polarization measurements add additional constraints to models~\cite{Lansberg:2011hi,Abelev:2011md,Aaij:2013nlm,Aaij:2014qea}. 
While the polarization of prompt $\jpsi$ measured by LHCb in the low-$\pt$ region is described by the NRQCD models, the predicted transverse polarization at high-$\pt$ is not observed in any of the LHC experiments~\cite{Aaij:2014qea}.

At the LHC energies, several parton interactions may occur in a single pp collision. Multi-parton interactions (MPI) influence the production of light quarks and gluons, affecting the total event multiplicity~\cite{Acosta:2004wqa,Khachatryan:2010pv,Aad:2010fh}. 
If the MPI also affect heavy-flavour production, this could introduce a dependence on the charged-particle multiplicity. 
A faster-than-linear correlation has been observed between the production of charged particles and that of prompt D mesons, as well as that of inclusive, prompt, and non-prompt $\jpsi$ in pp collisions at $\sqrt{s}=7$~TeV~\cite{Adam:2015ota}, suggesting that the heavy-flavour quark production mechanism is at the origin of this trend, while hadronization does not have a dominant effect. 
Recent results on inclusive $\jpsi$ (at midrapidity) in pp collisions at $\sqrt{s}=13$~TeV~\cite{Acharya:2020pit} confirm the observed correlation and extend the measurement multiplicity reach. 
In addition, measurements of the associated production of charmonia, bottomonia and/or open-charm hadrons in pp collisions at $\sqrt{s}=7$ and 8~TeV indicate that double-parton scatterings have a significant contribution to their production~\cite{Aaij:2012dz,Aaij:2015wpa}.

The structures seen in two-particle angular correlations in high-multiplicity pp and p--Pb collisions at the LHC are similar to those in Pb--Pb data~\cite{Khachatryan:2010gv,CMS:2012qk,Abelev:2012ola,Aad:2012gla,Aamodt:2011by}. These structures in Pb--Pb collisions are interpreted as signatures of the collective motion of particles in the quark--gluon plasma. In the heavy-flavour sector, the correlation of $\jpsi$ and charged particles in p--Pb collisions also shows similar features to those observed with charged hadrons~\cite{Acharya:2017tfn,Sirunyan:2018kiz}. The comparison with Pb--Pb measurements~\cite{Acharya:2017tgv,Acharya:2018pjd,Aaboud:2018ttm} suggests that there is a similar effect responsible for azimuthal asymmetries in both collision systems. The results at $\sqrt{s}=13$~TeV extend the charged-particle multiplicity reach up to about seven times minimum bias multiplicity, corresponding to about 50 charged particles per unit of rapidity, values similar to those of p--Pb collisions (about 40~\cite{ALICE:2018wma}) where collective-like effects have been observed.

In this publication, the measurement of the inclusive yield and the first moment of the \jpsi \pt distribution as a function of the charged-particle multiplicity in pp collisions at $\sqrt{s}=5.02$ and 13~TeV is presented. $\jpsi$ mesons are reconstructed via their dimuon decay channel at forward rapidity ($2.5<y<4$)\footnote{In the ALICE reference frame the muon spectrometer covers a negative pseudorapidity range. However, we use positive \y to represent symmetric collision systems (pp) considered in the present analysis.}, whereas the charged-particle multiplicity is measured at midrapidity ($|\eta|<1$). 
These measurements complement those performed for $\jpsi$ at forward rapidities in pp collisions at $\sqrt{s}=7$~TeV~\cite{Abelev:2012rz}.

\section{Experimental apparatus and data sample} 
\label{sec:evntsel}
The ALICE apparatus is described in detail in Refs.~\cite{Aamodt:2008zz,Abelev:2014ffa}. The detectors employed in this measurement, namely V0, Silicon Pixel Detector (\SPD), and Muon Spectrometer (MS), are described below.

The V0 detector consists of two scintillator hodoscopes located on each side of the interaction point $(2.8<\eta<5.1$ and $-3.7<\eta<-1.7)$~\cite{Abbas:2013taa}.
It provides a minimum bias (MB) trigger which requires a signal in both hodoscopes.
The charged-particle multiplicity at midrapidity is measured using the SPD~\cite{Aamodt:2010aa}. 
The SPD consists of two layers of silicon pixel detectors covering pseudorapidity ranges $|\eta|<2$ and $|\eta|<1.4$, respectively.
It is used to reconstruct the primary vertex as well as tracklets, short two-point track segments covering the pseudorapidity region $|\eta|<1.4$. Tracklets are required
to point to the primary interaction vertex within $\pm$1~cm in the transverse plane and $\pm$3~cm in the beam ($z$) direction. 

The muons originating from \jpsi decays are detected at forward rapidity ($2.5 < y < 4.0$) in the MS.
The MS is composed of five tracking stations with two planes of Cathode Pad Chambers for the first two stations and two planes of Cathode Strip Chambers for the rest. The third station is placed inside a dipole magnet with a field integral of 3 Tm. 
Two trigger stations, each with two planes of Resistive Plate Chambers, are positioned downstream of the tracking system and provide a single muon as well as a dimuon trigger.
A 4.1 m long front absorber of 10 interaction lengths ($\lambda_{\rm{int}}$) is placed between the interaction point and the first tracking station to stop the high hadron flux. Hadrons which
escape this front absorber are further filtered out by a second absorber, a  1.2 m long (7.2 $\lambda_{\rm{int}}$) thick iron wall, placed between the tracking and the triggering system, which also removes low-momentum muons originating from pion and kaon decays.
A conical absorber shields the muon spectrometer against the secondary particles produced by the interaction of primary particles in the beam pipe throughout the entire length of the MS.

The reported results are based on data collected in \pp collisions at $\s =$ 5.02 and 13~TeV by ALICE during 2015 and 2016, respectively.
The \jpsi production is measured using dimuon triggered events. 
A dimuon trigger is obtained as the coincidence of a MB trigger and at least one pair of track segments reconstructed in the muon trigger system with low \pt threshold of 0.5 \GeVc.
The analysis presented in this publication utilizes $\sim$1.2 million and $\sim$121.1 million  dimuon trigger events for \pp collisions at $\s =$ 5.02 and 13~TeV, respectively. %
This corresponds 
to an integrated luminosity of $\approx109.1$ ${\rm nb}^{-1}$ ( $\approx4.9$ ${\rm pb}^{-1}$) in \pp collisions at \s $=$ 5.02 (13)~TeV.
Events containing more than one distinct vertex were tagged as pileup and discarded from the analysis. The maximum probability of the pileup was about 5$\times$10$^{-3}$ for the data collected at $\s =$ 13 TeV, while at \s $=$ 5.02 TeV the pileup was below 2.5$\%$~\cite{Acharya:2017hjh}.

\section{Data analysis}

\jpsi yield as well as \dnchdeta are measured for INEL~$>$~0 events, which are defined as inelastic collisions for which at least one charged-particle track is recorded within $|\eta| <$ 1.0. 
Beam-induced background events are removed by using the time information from the V0 detectors and the correlation between the number of clusters and track segments reconstructed in the SPD. Pileup events, i.e. events with multiple collision vertices are removed as follows: using the V0 time information to select events within a bunch crossing time window (out-of-bunch pilepup); using a vertex finding algorithm based on the SPD~\cite{Adam:2015gka} to remove events with multiple vertices (in-bunch pileup).
The impact of a possible residual contamination from in-bunch pile-up was estimated by dividing the data sample in two groups, according to the average pileup probability per bunch crossing below/above 0.6\%. A negligible effect was observed on the final results by analysing the sub-sample below 0.6\% average
probability per bunch crossing.

\subsection{Charged-particle multiplicity density}
\label{sec:charged-particle}
The charged-particle multiplicity per unit of pseudorapidity (\dnchdeta) is estimated using information from the \SPD.  
A pair of hits in the two \SPD layers that are pointing towards the interaction vertex are chosen to form tracklets (track segments). The \dnchdeta  is proportional to the measured number of tracklets (\Ntrk). 
Primary charged particles, defined as all particles produced in the collision (including products of strong and electromagnetic decays) except those from the weak decay of strange hadrons~\cite{ALICE-PUBLIC-2017-005}, are considered for the analysis.

In order to minimize non-uniformities in the \SPD acceptance, only events with a $z$-vertex position within $|\zSPDv| < 10$ cm are considered,
and the tracklet multiplicity is measured within $|\eta| < 1$. 

In addition, 
the variation of the SPD efficiency and its number of inactive channels throughout the data taking period, causes the measured \Ntrk to vary with \zSPDv. About 4 to 15$\%$ dead channels were registered 
for the analyzed data sets. In order to correct for these detector effects, 
a data-driven correction is applied to equalize \Ntrk variation as a function of \zSPDv and time on an event-by-event basis~\cite{Abelev:2012rz,Adam:2015ota}. The \Ntrk for each event reconstructed for a given $z$-vertex position is corrected by the average fraction of missing tracklets at the interaction $z$-vertex with respect to a reference multiplicity. 
The maximum of the average \SPD tracklet multiplicity as a function of \zSPDv position was chosen as the reference value or $\meanNtrkcorr(\zSPDv)$ profile, similarly as done in previous analyses~\cite{Abelev:2012rz,Adam:2015ota}. However, minimum reference multiplicity has also been tested and the result was found to be consistent with the maximum one.  
The correction factors are smeared randomly using a Poisson (or Binomial depending on the reference value) distribution~\cite{Adamova:2017uhu,Abelev:2012rz}. 

Monte Carlo (MC) simulations based on  PYTHIA6~\cite{Sjostrand:1993yb,Sjostrand:2006za}, PYTHIA8~\cite{Sjostrand:2014zea} and EPOS-LHC~\cite{Pierog:2013ria}  event generators and GEANT3~\cite{Brun:1082634} transport code were used for evaluating the correlation between \Nch and \Ntrkcorr.
Assuming a second order polynomial function ($f$) to describe the correlation between \Ntrkcorr and \Nch, the average charged-particle multiplicity density in a given multiplicity interval $i$, is estimated as~\cite{Acharya:2020giw}
\begin{equation}
\frac{\left ({\rm d}N_{\rm ch}/{\rm d}\eta \right )^{i}}{\langle {\rm d}N_{\rm ch}/{\rm d}\eta \rangle} = \frac {f( \langle \Ntrkcorr \rangle^{i})} {\Delta\eta \times \langle {\rm d}N_{\rm ch}/{\rm d}\eta \rangle}.
\end{equation}
where $\Delta\eta = 2$ is the pseudorapidity window in which charged particles are measured.
Possible deviations from this assumption are evaluated by either assuming a linear correlation between \Ntrkcorr and \Nch or using a Bayesian unfolding procedure~\cite{Adye:2011gm,DAgostini:1994fjx}.
The values of \Nch obtained using these methods are consistent within systematic uncertainties.
The $\langle \dnchdeta \rangle$ is measured by ALICE in the same acceptance and is found to be 5.48 $\pm$ 0.05 (uncorr. syst.) $\pm$ 0.05 (corr. syst.) and 6.93 $\pm$ 0.07 (uncorr. syst.) $\pm$ 0.06 (corr. syst.) for pp collisions at \s $=$ 5.02 TeV and 13 TeV, respectively~\cite{Acharya:2020kyh}. 
A summary of \avgdnchdeta values for the $\rm INEL > 0$ class is shown in Table~\ref{tab:ppdnch}. Statistical uncertainties are negligible.

\begin{table}[htb]
\centering
\caption{The corrected tracklet \Ntrkcorr intervals and the corresponding relative charged-particle multiplicity densities \avgdnchdeta along with their systematic uncertainties are listed for \pp collisions at \s $=$ 5.02 and 13 TeV. 
}
\begin {tabular}{c c | c c}
\hline
\multicolumn{2}{c}{ \s $=$ 5.02 TeV } & \multicolumn{2}{c}{\s $=$ 13 TeV} \\\hline
\Ntrkcorr & $ \avgdnchdeta $ & \Ntrkcorr & $\avgdnchdeta $ \\ 
\hline
 $1-7$     & 0.44$\pm$0.01            & $1-8$     &0.38$\pm$0.01\\
 $8-12$    & 1.11$\pm$0.01            & $9-14$    &1.01$\pm$0.01\\
 $13-18$   & 1.70$\pm$0.02            & $15-20$   &1.54$\pm$0.02\\
 $19-29$   & 2.51$\pm$0.04            & $21-25$   &2.03$\pm$0.03\\
 $30-48$   & 3.80$\pm$0.07            & $26-33$   &2.56$\pm$0.04\\
 $49-100$  & 5.71$\pm$0.16            & $34-41$   &3.23$\pm$0.05\\
 $ $         & $ $                    & $42-50$   &3.92$\pm$0.07\\
 $ $         & $ $                    & $51-60$   &4.68$\pm$0.09\\
 $ $         & $ $                    & $61-80$   &5.67$\pm$0.12\\
 $ $         & $ $                    & $81-115$  &7.28$\pm$0.19\\
 \hline
\end {tabular}
\label{tab:ppdnch}
\end {table}  

\subsection{ \jpsi reconstruction and signal extraction}
\label{sec:signal}
In order to provide muon identification, each candidate track reconstructed in the muon tracking chamber (MCH) of the MS is required to match with a corresponding track segment in the muon trigger (MTR).  
A selection on the pseudorapidity ($-4< \eta <-2.5$) is applied on both J/$\psi$ daughter tracks to reject muons at the edges of the spectrometer
acceptance.
In addition, a selection on the distance from the $z$ axis  of the track at the end of the absorber within 17.6 to 89.5 cm is applied. 
The selection removes tracks which cross the high density part of the absorber, since they are significantly affected by multiple Coulomb scattering, which results in poor mass resolution of the corresponding dimuon pair.
A condition on the rapidity of the dimuon pair, $2.5< \y <4$, is applied. 

The \jpsi raw signal yield is obtained by fitting the opposite sign dimuon invariant mass distribution with a superposition of \jpsi and $\psi(2S)$ signals and a function to account for the background. The estimation of the \jpsi signal and background is carried out using several fit functions. The \jpsi signal is extracted using two functions, an extended Crystal Ball function, consisting of Gaussian core and asymmetric power-law tails, and a pseudo-Gaussian with mass-dependent asymmetric tails~\cite{ALICE-PUBLIC-2015-006}. The \jpsi peak position and width are left free in the fit, while the tail parameters are fixed to the values obtained from MC simulations anchored to the data with injected \jpsi signals. Similarly, as a part of the checks related to the systematic uncertainty of the signal extraction, tail parameters from a previous analysis at \s $=$ 13 TeV are also used~\cite{Acharya:2017hjh}.
The mass position and width of the $\psi(2\rm S)$ are bound to the mass and width of \jpsi using the method described in Ref.~\cite{Acharya:2017hjh}. 
Similar approaches are considered for the signal extraction at $\sqrt{s}$ $=$  5.02 and 13 TeV, except the different mass ranges and background functions. 
The invariant mass fit is performed within [2, 5] GeV/$c^{2}$ at both energies, and this range was varied in order to compute the systematic uncertainty.

\begin{figure}[h!]
\subfigure{\includegraphics[scale=0.41]{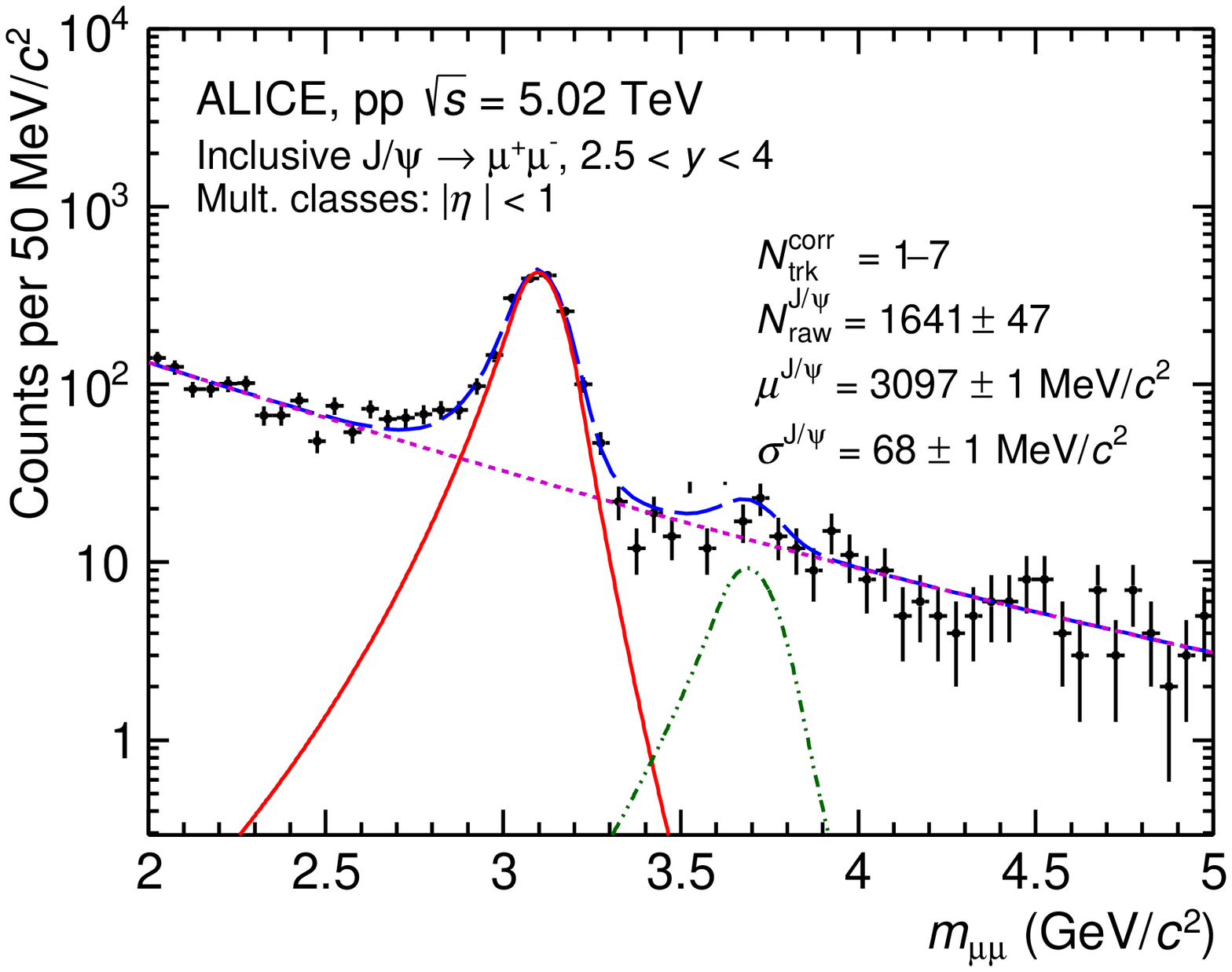}}
\subfigure{\includegraphics[scale=0.41]{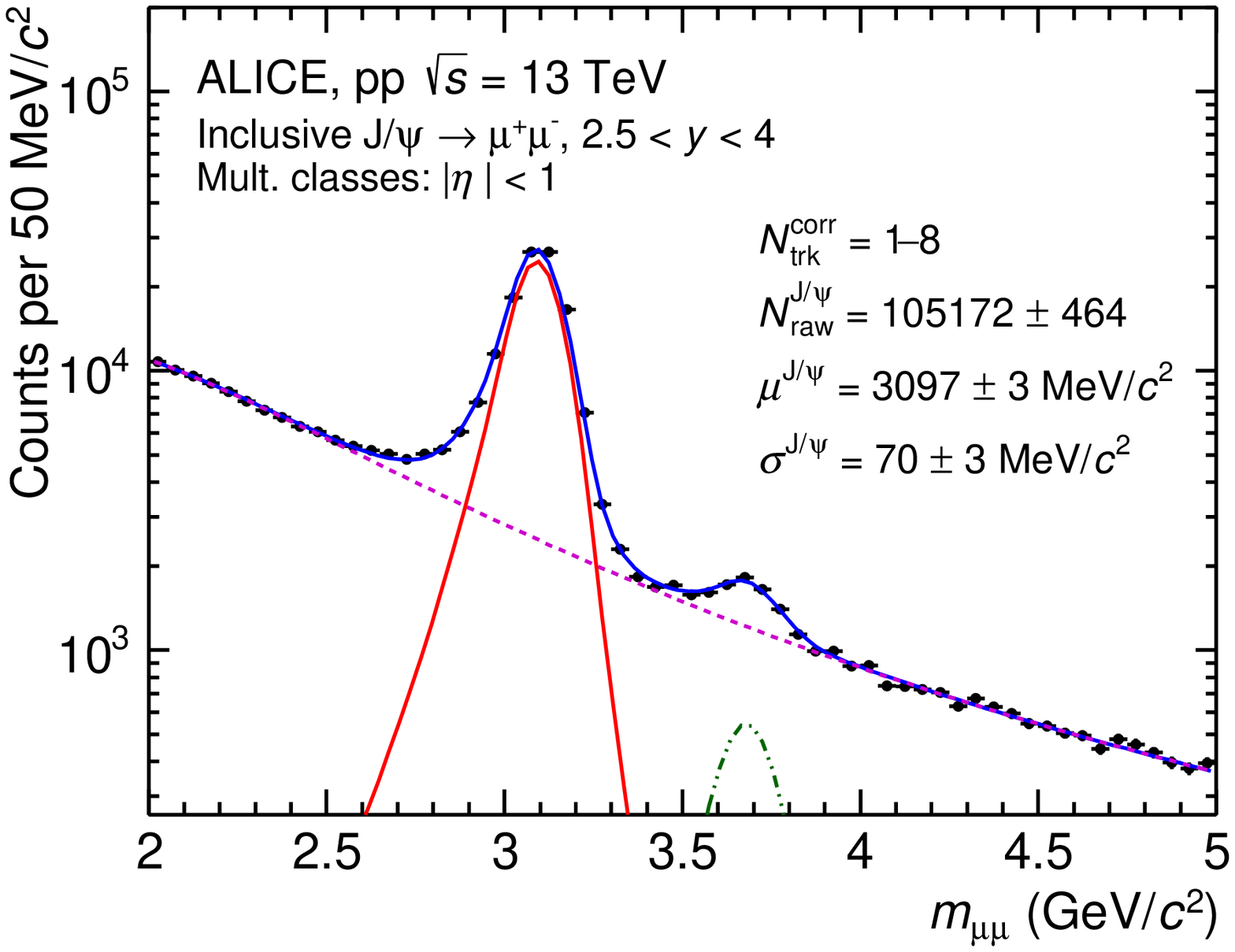}}
\subfigure{\includegraphics[scale=0.41]{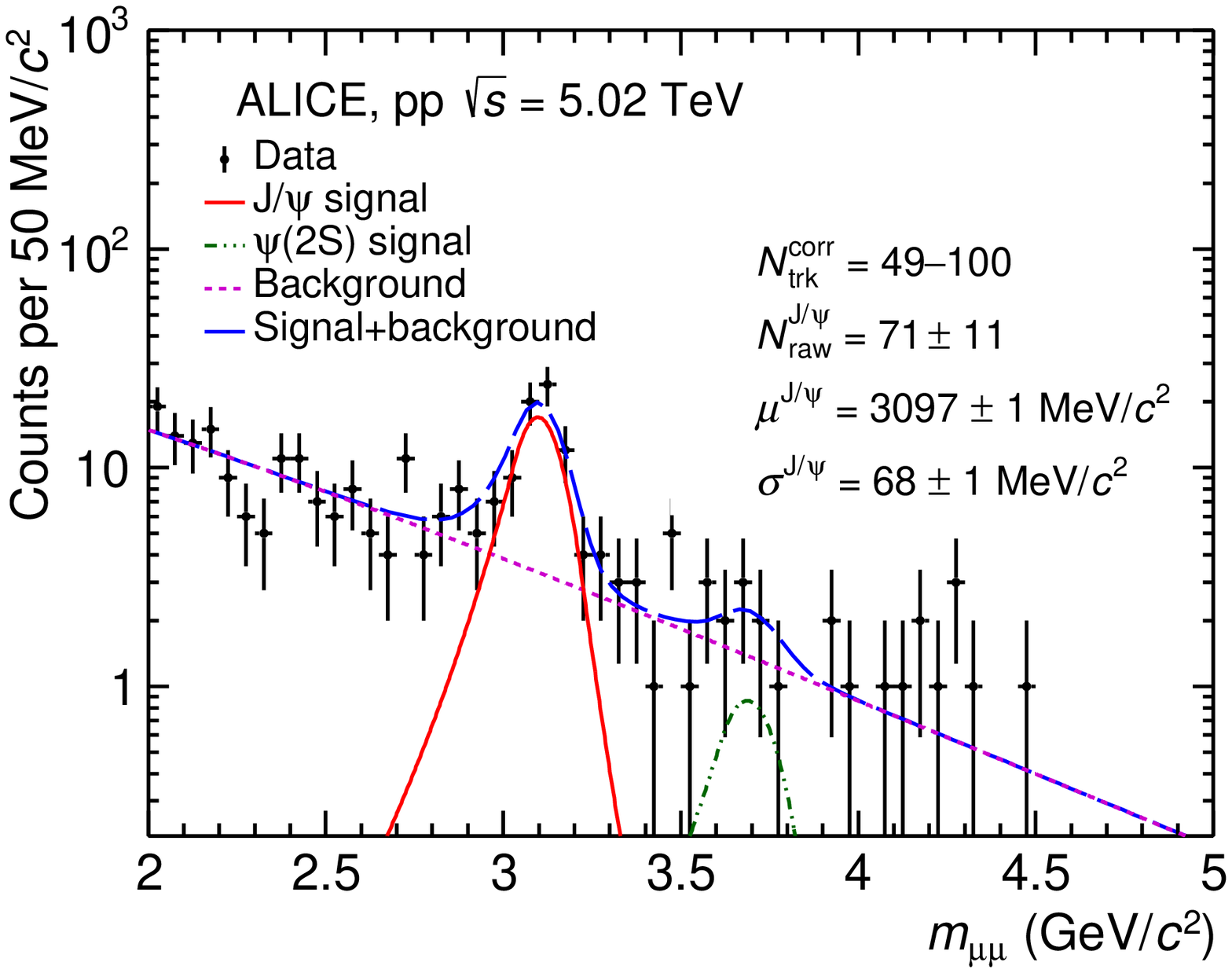}}
\subfigure{\includegraphics[scale=0.41]{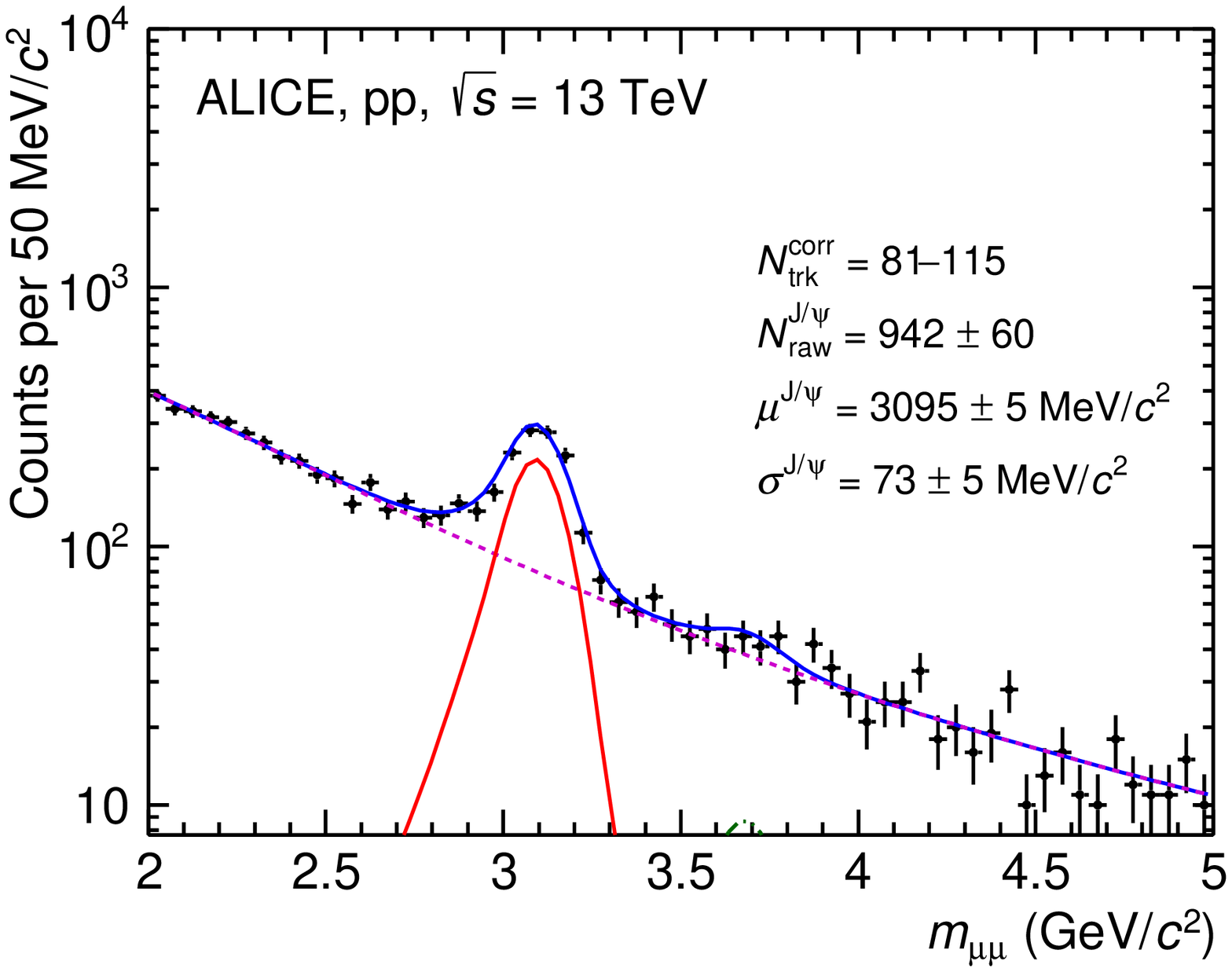}}
\caption{\footnotesize{Dimuon invariant mass distribution in the lowest (top panels) and highest (bottom panels) \Ntrkcorr intervals in \pp collisions at $\sqrt{s} =$ 5.02 (left) and $\sqrt{s} = 13$ TeV (right). Example fit functions are shown by lines for \jpsi, $\psi(2\rm S)$ signal, background and the combination of signal with background. The data points with error bars are represented by scattered points.}}
\label{fig:jpsisignal}
\end{figure}

At \s $=$ 5.02 TeV, the background is parameterized with either a Gaussian with a width that varies linearly with the mass, called Variable Width Gaussian function, or a ratio of a first-order to a second-order polynomial function.
In the case of \s $=$ 13 TeV, the background functions are the Variable Width Gaussian function, and the product of a fourth order polynomial and an exponential function. Examples of \jpsi signal extraction in the lowest and highest multiplicity intervals at both the collision energies are shown in Fig.~\ref{fig:jpsisignal}. 
Several tests are performed by taking different combinations of signal and background functions, invariant mass ranges as well as tail parameters. 
The raw yields of \jpsi and the corresponding statistical uncertainties are calculated by taking the mean of the values obtained in these tests. The square root of variance (RMS) of different tests is assigned as systematic uncertainty.
The dimuon invariant-mass distribution is sliced in various \Ntrkcorr intervals as discussed in Section~\ref{sec:charged-particle}.
The signal extraction in each multiplicity interval is performed in the same way as in the multiplicity-integrated case. 

The \jpsi yield is extracted from the uncorrected invariant mass distribution and corrected for \jpsi acceptance and efficiency (\acceff) of the MS.
The \acceff factor is estimated based on MC simulations. A \jpsi (\pt,\y) distribution corresponding to the data is used in the simulations. The generated (\pt,\y) distributions are modified via an iterative process to reproduce the reconstructed ones. The MC simulation is performed enforcing \jpsi decays into two opposite-sign muons along with radiative photon emissions using EVTGEN~\cite{Lange:2001uf} and PHOTOS~\cite{Barberio:1990ms}. The simulated decay muons are transported through the detector using GEANT3~\cite{Brun:1082634}.

\subsection{Relative \jpsi yield measurement}
\label{sec:ryield}
The relative yield of \jpsi in a given multiplicity interval $i$ is computed by normalizing to the corresponding multiplicity integrated value as

\begin{equation}
\centering
 \frac{\left( {\rm d}N^{\jpsi}/{\rm d}y \right)^{\rm i}}{\langle {\rm d}N^{\jpsi}/{\rm d}y \rangle }  = \frac{N^{\jpsi,\rm i}}{N^{\jpsi}}\times\frac{N^{\rm MB}_{\rm {eq}}}{N^{\rm {MB,i}}_{\rm {eq}}}\times\frac{\left(\acceff\right)^{\jpsi}}{\left(\acceff\right)^{\jpsi,\rm i}}\times\frac{{\epsilon}^{\rm MB,i}}{{\epsilon}^{\rm MB}}\times\frac{\epsilon^{\jpsi}}{\epsilon^{\jpsi,\rm i}},
\label{eq:relY3}
\end{equation} 

where $N^{\jpsi}$ and $N^{\rm MB}_{\rm eq}$ are the number of \jpsi reconstructed candidates and the equivalent number of MB events. 
$N^{\rm MB}_{\rm eq}$ is obtained from the number of dimuon triggered events, $N_{\rm \mu\mu}$, as $N^{\rm MB}_{\rm eq} = F_{\rm Norm} \times N_{\rm \mu\mu}$. The normalization factor $F_{\rm Norm}$ (see Sec.\ref{sec:systematic}) was estimated
as in previous analyses~\cite{Adamova:2017uhu,Acharya:2020giw}. The \acceff correction for $N^{\jpsi}$ is taken into account for the multiplicity integrated case, $\left(\acceff\right)^{\jpsi}$, as well as in multiplicity bins, $\left(\acceff\right)^{\jpsi,\rm i}$. The \acceff is observed to increase by around 4\% from low to high multiplicity at both center-of-mass energies. This is because \acceff(\pt,$y$) increases with \pt and that \meanpt increase with multiplicity.

The $1/\epsilon^{\rm MB}$ and $1/\epsilon^{\jpsi}$ represent correction factors applied on the number of MB selected events and number of \jpsi candidates, respectively. These factors account for possible event and signal losses due to the event selections applied at the analysis level. Each of these corrections include contributions from the minimum bias trigger efficiency for INEL $>$ 0 selection ($\epsilon^{\rm MB}_{\rm INEL>0}$), $z$-vertex range ($\epsilon^{\rm MB}_{\rm vtx,range}$), vertex quality selection ($\epsilon^{\rm MB}_{\rm vtx,QA}$), and pileup rejection ($\epsilon_{\rm pu}$). 
The correction factors for MB event selection and \jpsi are studied as a function of the event multiplicity. It is observed that the corresponding values are close to unity in all multiplicity event classes, except for the lowest multiplicity interval where the total factors become 0.89 and 0.92 at \s $=$ 5.02 and 13 TeV, respectively (Table~\ref{tab:ppeffi}). 
The correction to MB trigger event selection for the $\rm INEL>0$ class was estimated from MC by evaluating the fraction of $\rm INEL>0$ events that satisfy the MB trigger condition.
The correction factors for measuring \jpsi(and MB events) that assure good vertex quality within the satisfied trigger condition were estimated from data by taking the ratio between the number of \jpsi(and MB triggered events) with and without vertex selection criteria. 
The effect of the vertex range ($|\zSPDv| <$ 10 cm) selection is the same for MB and \jpsi, hence this efficiency correction will cancel out in the ratio. The correction for the events that are rejected due to the pileup is evaluated from data by comparing $N^{\rm MB}_{\rm i}/\langle N^{\rm MB}\rangle$ between the high and low pileup rate runs in each multiplicity interval with and without pileup selection criteria. A similar technique was applied for \jpsi as well, using dimuon events.
In addition, the integrated number of MB events is used to normalize the \jpsi yield, which includes events with zero number of charged particles ($\rm INEL =$ 0 events). This contamination is taken into account by a correction factor ($1/\epsilon_{\rm INEL = 0}$) estimated through MC simulations.
The values of all efficiency correction factors for the integrated case over multiplicity, as well as for the lowest multiplicity interval, are summarized in Table~\ref{tab:ppeffi}. 
\begin{table}[htb]
\footnotesize
\centering
\caption{The various efficiency factors which are applied to calculate the relative yield of \jpsi along with their statistical uncertainty. The values quoted without uncertainty have negligible statistical uncertainty. }
\begin {tabular}{c |c | c }
\hline
Efficiency & \s $=$ 5.02 TeV & \s $=$ 13 TeV \\[0.2cm] 
\hline 
$\epsilon^{\rm MB}_{\rm INEL>0}$ & 93$\%$  & 94$\%$  \\ [0.1cm] \hline
$\epsilon^{\rm MB}_{\rm INEL>0}$ (lowest interval)& 89$\%$ & 92$\%$ \\ [0.1cm]\hline 
$\epsilon^{\rm MB}_{\rm vtx,QA}$ & 96$\%$ & 94$\%$ \\ [0.1cm]\hline
$\epsilon^{\jpsi}_{\rm vtx,QA}$ & 99.0$\pm$0.2$\%$ & 97.2$\pm$0.3$\%$ \\[0.1cm] \hline
$\epsilon_{\rm INEL=0}$ & 95$\%$ & 98$\%$ \\[0.1cm] \hline
\end {tabular}
\label{tab:ppeffi}
\end {table}    

\subsection{Average transverse momentum, \meanptjpsi measurement}
\label{sec:meanpt}
The average transverse momentum of \jpsi (\meanptjpsi), is extracted by fitting the acceptance times efficiency corrected dimuon mean transverse momentum (\ptmu) as a function of invariant mass of the dimuon ($m_{\mu^{+}\mu^{-}}$). This method allows us to extract the \meanptjpsi in multiplicity intervals with a low number of reconstructed \jpsi mesons.
A two dimensional map of acceptance and efficiency, \acceff(\pt,\y) is produced using simulated events, as described in Section~\ref{sec:signal}, to correct the \ptmu distributions. The \acceff corrected \ptmu distribution is studied as a function of the charged-particle multiplicity. A phenomenological function is used to extract the signal. It is based on weighting the \meanpt of J/$\psi$ entering in the spectrum and the \meanpt of the background, by the ratio of the signal over signal plus background of each particle ( J/$\psi$ and $\psi$ (2S) ). Similar technique was previously used in Refs.~\cite{Adamova:2017uhu,Acharya:2020giw}.

\begin{figure}[tb]
\subfigure{\includegraphics[scale=0.83]{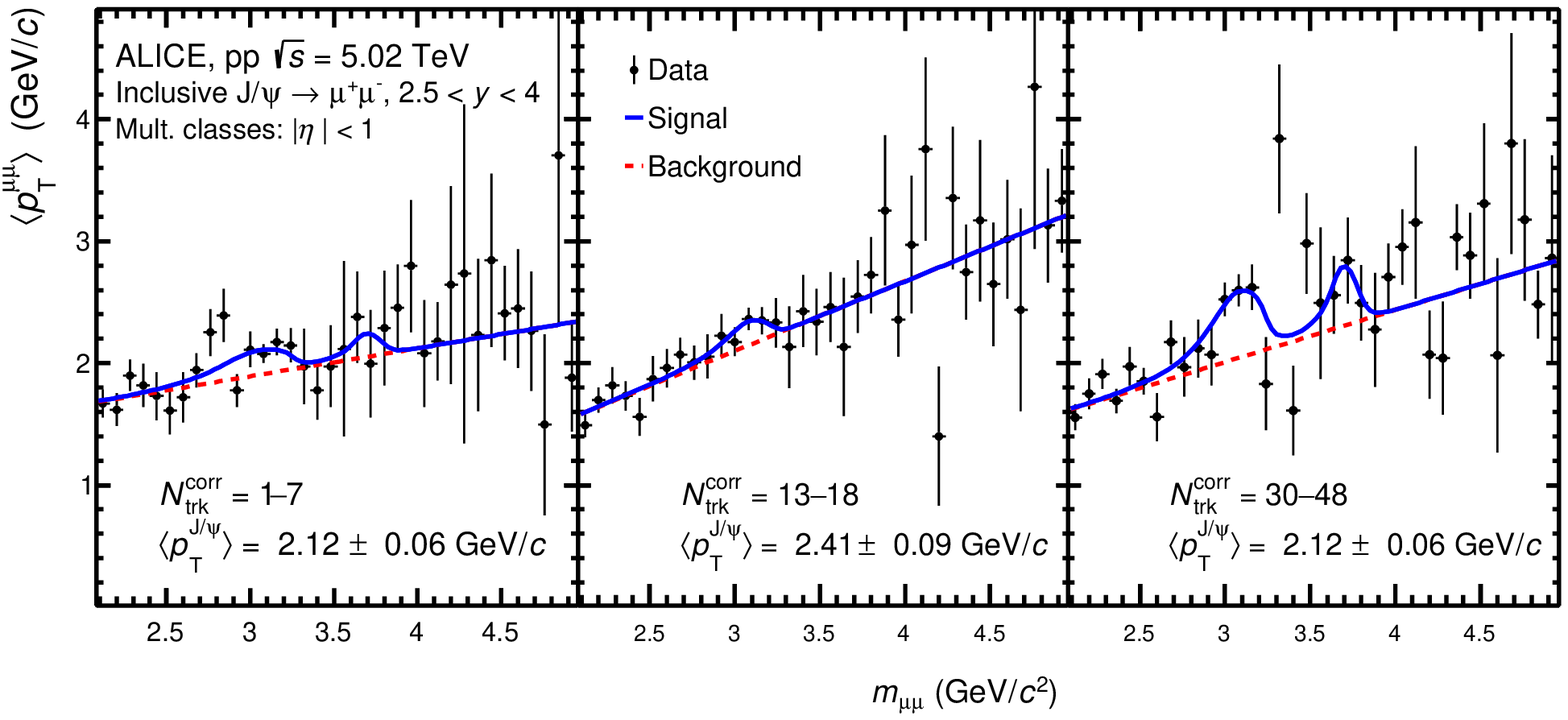}}
\subfigure{\includegraphics[scale=0.83]{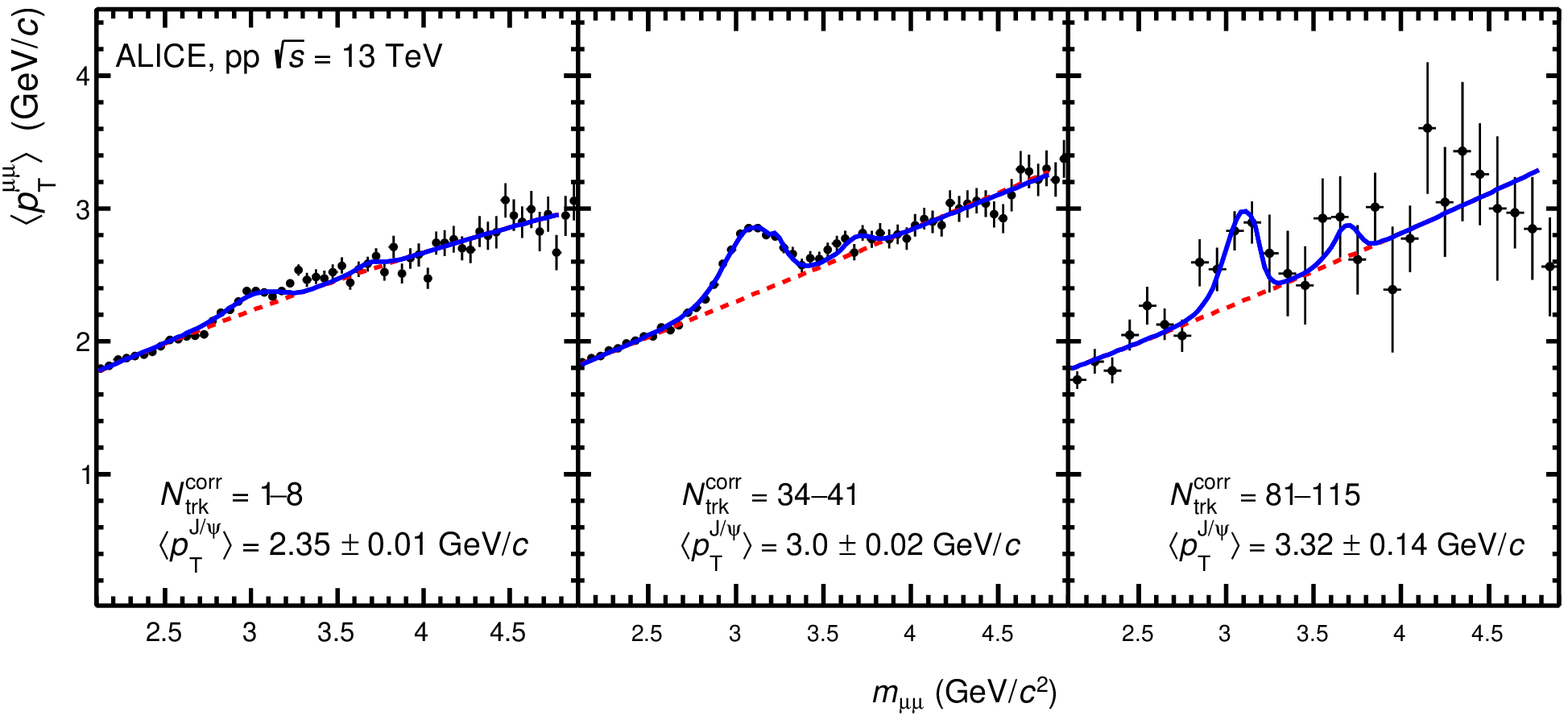}}
\caption{Example of $\langle p^{J/\psi}_{\it{T}}\rangle$ extraction in highest, intermediate and lowest \Ntrkcorr bins in \pp collisions at $\sqrt{s} =$ 5.02 and 13 TeV. The scattered points represent the experimental data while various lines show the signal and background functions.}
\label{fig:meanptsignal}
\end{figure}

During the \meanptjpsi extraction procedure, the ratios of signal over the sum of signal and background of the two charmonium states are fixed to the value extracted from fitting the \acceff corrected dimuon invariant-mass spectrum.
Three sets of alternative background functions are used to measure the \meanptjpsi.
The invariant mass ranges are the same as for \jpsi yield measurements. Fit examples of \ptmu distributions used to extract \meanptjpsi are shown in Fig.~\ref{fig:meanptsignal} for the highest, lowest, and intermediate multiplicity intervals.

\begin{figure}[tb]
\subfigure{\includegraphics[scale=0.415]{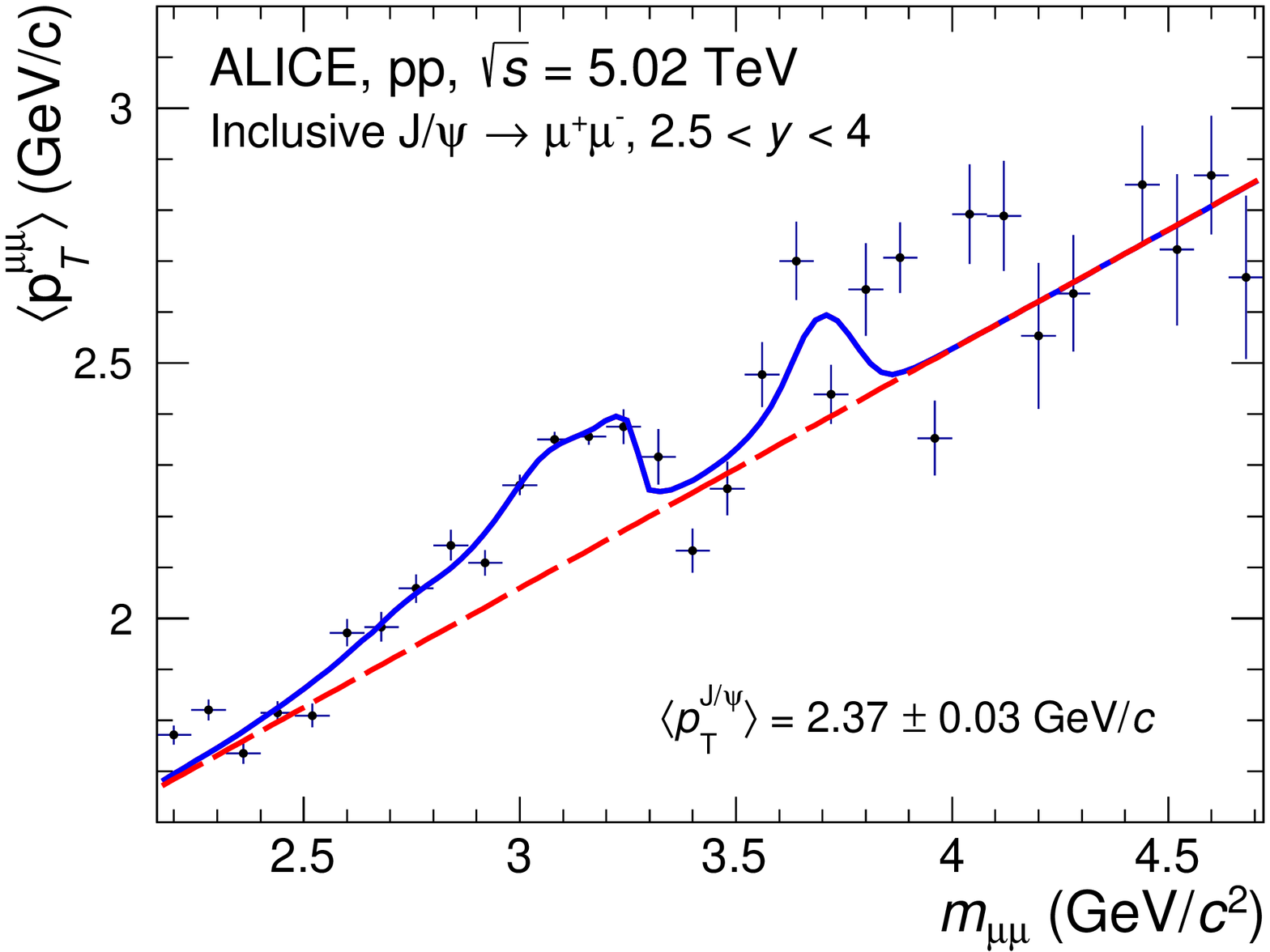}}
\subfigure{\includegraphics[scale=0.425]{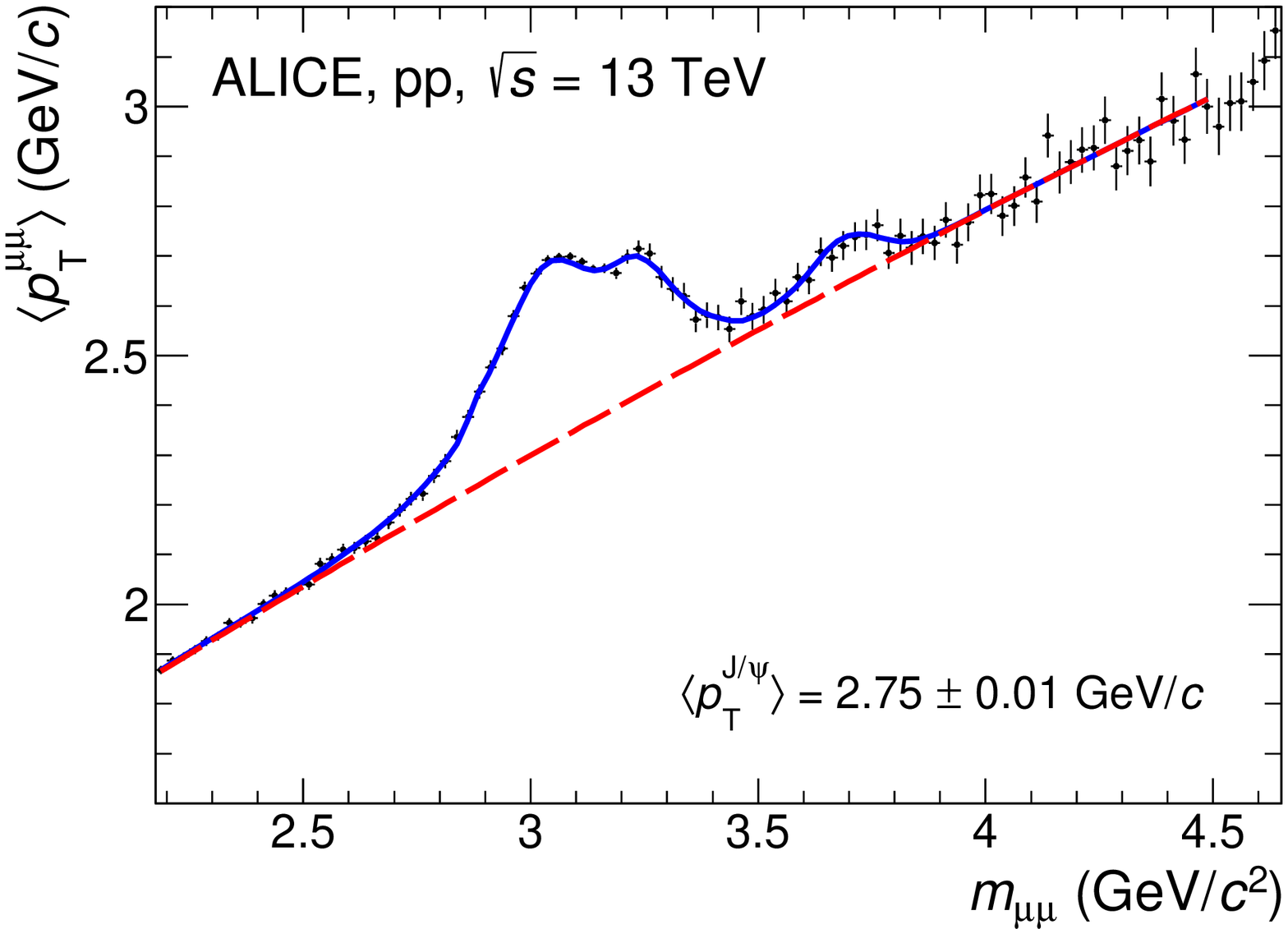}} 
\caption{Example of $\langle p^{J/\psi}_{\it{T}}\rangle$ extraction in integrated multiplicity using the piece-wise function for \pp collisions at $\sqrt{s} = 5.02$ and 13 TeV.}
\label{fig:Intgmeanpt}
\end{figure}

 A second peak around the signal region, already observed in previous analyses~\cite{Adamova:2017uhu}, can be seen for the multiplicity integrated case, as shown in Fig.~\ref{fig:Intgmeanpt}, which is due to the fact that the \meanptjpsi is not constant with mass, rather it is a function of $m_{\rm{\mu^{+}\mu^{-}}}$. It is worth noticing that the dimuon invariant mass resolution depends on the precision in the measurement of the momentum of each muon and the angle between them. Several effects like scattering and energy-loss fluctuations in the muon absorber, and the precision of the tracking chambers may affect the precision of the measurements. These effects may induce a variation of the
 \ptmu as a function of the  $m_{\rm{\mu^{+}\mu^{-}}}$. The variation of \meanptjpsi with dimuon invariant mass was quantified through MC simulations using pure signal, in particular a closure test was performed and it is found to be successful when the ``piece-wise'' function is used to parametrize this variation. This function is given by,

\begin{equation}
h(m_{\mu^{+}\mu^{-}}) = a_{0} + \sum^{5}_{i=1} a_{i}.(m_{\mu^{+}\mu^{-}} -  m_{J/\psi}),
\end{equation}

where,
\begin{equation}
\langle p^{J/\psi}_{T} \rangle(m_{\mu^{+}\mu^{-}}) = 
\begin{cases}
h(r_{0}) + a_{6}.( m_{\mu^{+} \mu^{-}} - r_{0})  & m_{\mu^{+} \mu^{-}} \leq  r_{0}\\
h(m_{\mu^{+} \mu^{-}})                     & r_{0} <  m_{\mu^{+} \mu^{-}} <  r_{1} \\
h(r_{1}) + a_{7}.( m_{\mu^{+} \mu^{-}} - r_{1})    & m_{\mu^{+} \mu^{-}} \geq r_{1}.\\
\end{cases}
\label{eq:piecewise}
\end{equation}

The fit function is evaluated at $m_{\rm{\mu^{+}\mu^{-}}} = m_{\rm{J}/\psi}$, and $a_{0}$ gives \meanptjpsi. Some of the parameters of the fit function are fixed from MC simulation by considering the reconstructed $\langle p_{\rm T} \rangle$ as a function of invariant mass. The piece-wise fitting function is able to recover the true \meanpt in a closure test performed using MC simulations. As shown in Fig.~\ref{fig:Intgmeanpt}, this function reproduces well the variation of the dimuon $\langle p_{\rm T} \rangle$ in the invariant mass boundary limit of the piece-wise function denoted by $r_{0}$ and $r_{1}$. The fits are performed considering $r_{0} =$ 2.8 GeV/$c^{2}$ and $r_{1} =$ 3.2 GeV/$c^{2}$ ( $r_{0} =$ 2.9 GeV/$c^{2}$ and $r_{1} =$ 3.4 GeV/$c^{2}$ ) at \s $=$ 5.02 (13) TeV. The boundary limits are determined from MC simulations.~The central value of \meanptjpsi and its uncertainties are calculated similarly as for the \jpsi yield analysis. The values of \meanptjpsi are measured to be 2.37 $\pm$ 0.03 (stat.) $\pm$ 0.01 (syst.) for multiplicity integrated events at $\sqrt{s} = $ 5.02 TeV and 2.76 $\pm$ 0.01 (stat.) $\pm$ 0.01 (syst.) at $\sqrt{s} =$ 13 TeV, which are consistent within uncertainties with previous ALICE measurements which use the traditional method of extracting \meanptjpsi from \pt spectra~\cite{Acharya:2017hjh}.

\subsection{ Evaluation of systematic uncertainties}
\label{sec:systematic}
This section presents the summary of the assessment and implementation of the systematic uncertainties for the observables discussed in this analysis.  
The \Nch is calculated using different MC event generators as mentioned in Section~\ref{sec:charged-particle}. The variations of $\langle \Nch \rangle$ from the various MC event generators are taken as a contribution to the systematic uncertainty on $\langle \Nch \rangle$. 
The dependence of $\langle \Nch \rangle$ on the reconstructed \zSPDv is considered to estimate another contribution to the systematic uncertainty. 
In particular, the correlation between \Nch and \Ntrkcorr is studied in various \zSPDv intervals and the differences are taken as systematic uncertainty for \dnchdeta. The variation of \meanNtrkcorr as a function of \zSPDv for reconstructed MC events is not identical to that in data. Hence, the $\meanNtrkcorr(\zSPDv)$ profile from both data and MC are used for the correction of tracklets using a data driven method. The deviation across data and MC profiles acts as another source for estimating the systematic uncertainty on \Nch. All the mentioned sources are taken into account to evaluate \Nch. The average over all the combinations is considered as central value, and the root mean square deviation from this value as the systematic uncertainty. The total uncertainty on \Nch ranges within 0.5\%$-$2.4\% and 0.3\%$-$2.3\% for $\sqrt{s} = 5.02$ and 13 TeV, respectively. 
The different contributions of systematic uncertainties on \Nch are presented in Table~\ref{tab:systfinalNch}.
The MB trigger efficiency associated to the condition $\rm INEL>0$ enters in the estimate of $\langle\Nch\rangle$ in the lowest multiplicity interval and the systematic uncertainty assigned due to the correction is listed in Table~\ref{tab:systfinalNch} for both collision energies. The systematic uncertainty of the integrated \dnchdeta is taken from Ref.~\cite{Acharya:2020kyh}.
 
The systematic uncertainty due to signal extraction is determined by the RMS value of the $N^{\jpsi}_{\rm i}/N^{\jpsi}_{\rm tot}$ distribution, and represents the dominant source of uncertainty on the relative \jpsi yields, as shown in Table~\ref{tab:systfinal}. The relative ratios of multiplicity dependent \jpsi yields to the multiplicity integrated one are estimated by considering different fit mass ranges, tail parameters and background functions.

$F_{\rm Norm}$ is computed using different methods, for the integrated multiplicity as well as in the \Ntrkcorr intervals~\cite{Acharya:2020giw}.
In the first method, $F_{\rm Norm}$ is obtained as the ratio of the number of MB-triggered events to the number of MB-triggered events that also satisfy the dimuon trigger condition.
The second method benefits from the probability of occurrence of the single-muon triggered events by looking at the product of the probability of finding dimuon events in the single-muon triggered events and the probability of coincidence of single-muon and MB-triggered events (two-step offline method). The third method is based on the information regarding the dimuon and online MB trigger counters. Alternatively, $F_{\rm Norm}$ is also obtained by re-scaling the two-step offline method with the ratio of $N^{\rm MB}_{\rm i}/N^{\rm MB}$ to the $N^{\rm i}_{\rm \mu\mu}/N_{\rm \mu\mu}$. 
Run-by-run variations of $F^{i}_{\rm Norm}/F_{\rm Norm}$ values computed for the mentioned methods produce a maximum variation of 1.5$\%$ assigned as systematic uncertainty
at \s = 5.02 TeV, which is found to be independent of the multiplicity.

The influence of various efficiency correction factors on the relative yield of \jpsi are described in Section~\ref{sec:ryield}. Different MC generators are used to evaluate the efficiency factors and the difference across MC generators is assumed as the systematic uncertainty. The corresponding systematic uncertainties for $\epsilon^{\rm MB}_{\rm INEL>0}$ and $\epsilon^{\rm MB}_{\rm vtx,QA}$ are 1$\%$ and 2$\%$, respectively, for pp at \s $=$ 5.02 TeV. However, the systematic uncertainty for the same correction factor in pp at \s $=$ 13 TeV can be safely ignored as the difference in obtained values between PYTHIA8 and EPOS generators is negligible. The systematic uncertainties for $\epsilon_{\rm INEL=0}$ and $\epsilon^{\rm MB}_{\rm INEL>0}$ in the lowest interval are 1$\%$ (1.5$\%$) and 1.4 $\%$ (0.6$\%$) at \s $=$ 5.02 (13) TeV, respectively. Table~\ref{tab:systfinal} shows the systematic uncertainty of each correction factor. 
The systematic uncertainty associated with $\epsilon_{\rm pu}$ is estimated from MC and found to be negligible at both the collision energies. The discussed systematic uncertainties are considered uncorrelated.

\begin{table}[h!!!]
\footnotesize
\centering
\caption{The summary of systematic uncertainties for charged-particle density measurements. The value marked with (*) is taken from the Ref.~\cite{Acharya:2020kyh}.}
\begin {tabular}{c|c|c}
\hline
Source                                   &$\sqrt{s} = 5.02$ TeV  & $\sqrt{s} = 13$ TeV\\
\hline 
MC event generator                                  & $<1.0\% $ & $<1.0\%$ \\ 

$z_{\rm vtx}$ dependence                          & $0.2\% - 2.2\%$ & $0.2\% - 2.2\%$ \\

$z_{\rm vtx}~vs.~\langle \Ntrkcorr \rangle$  profile         & $<1.0\% $& $ 0.0-1.5\%$\\

$\langle \Nch \rangle$ (lowest interval)               & 1.3\%  & 0.3\% \\

$\langle\dnchdeta\rangle*$                                       & $1.3\%$& 1.3\%\\
\hline
\end {tabular}
\label{tab:systfinalNch}
\end{table}

\begin{table}[h!!!]
\footnotesize
\centering
\caption{The summary of systematic uncertainties for relative \jpsi yield measurements.~~~~~~~~~~~~~~~~~~~~~~~~~~~~~~~~~~~~~~~~~~~~~~~~~~~~~}
\begin {tabular}{c|c|c}
\hline
Source                                   &$\sqrt{s} = 5.02$  & $\sqrt{s} = 13$\\
\hline 
Signal\ extraction                          & $0.5\% - 2.8\%$  & $3.5\% - 5.9\%$ \\

F$_{\rm{norm}}$                                 & $0.2\% - 1.5\%$  & negligible \\
$\epsilon^{\rm MB}_{\rm INEL>0}$                        & $1.0\%$ & negligible \\

$\epsilon^{\rm MB}_{\rm vtx,QA}$                          & $2.0\%$ & negligible \\

$\epsilon^{\rm MB}_{\rm INEL>0}$ (lowest interval)          & $ 1.4\% $ & $0.6\% $\\

$\epsilon_{\rm INEL=0}$                                       & 1.0$\%$  & 1.5$\%$ \\

Pileup                                              & negligible & negligible \\

\hline
\end {tabular}
\label{tab:systfinal}
\end {table}

\begin{table}[h!!!]
\footnotesize
\centering
\caption{ The summary of systematic uncertainties for relative $\langle p^{J/\psi}_{\it{T}} \rangle$ measurements. Values marked with (*) are taken from the Ref.~\cite{Acharya:2017hjh}.}
\begin {tabular}{ c|c c|c c}
\hline
                   &  \multicolumn{2}{c}{ \s $=$ 5.02 TeV } & \multicolumn{2}{c}{\s $=$ 13 TeV} \\\hline
Source             & \meanptjpsi & \meanptjpsi/\meanptjpsi$^{\rm int}$ & \meanptjpsi & \meanptjpsi/\meanptjpsi$^{\rm int}$\\
\hline 
Signal~extraction       & 0.4 -- 4.1$\%$ & 0.5 -- 4.1$\%$ &0.1 -- 4.8$\%$ & 0.2 -- 4.9$\%$  \\
Acc $\times$ Eff       & 0 -- 2.1$\%$ & 0 -- 2.1$\%$ & 0 -- 3.0$\%$ &  0 -- 3.0$\%$  \\
Mass variation              & $0.4\%$ & -- & $0.7\%$ & -- \\
MC input mismatch           & $1.5\%$ & -- & $2.4\%$ & -- \\
MCH efficiency*             & $1.0\%$ & -- & $1.0\%$ & --\\
MTR efficiency*             & 2.0$\%$& -- & 4.0$\%$ & -- \\
Matching efficiency*                   & $1.0\%$ & -- & $1.0\%$ & --\\
\hline
\end {tabular}
\label{tab:systmeanpt}
\end {table}

In the \meanptjpsi analysis, the systematic uncertainty related to the signal extraction is estimated by taking the influence of various background fit functions, tail parameters, and invariant mass ranges. The uncertainty values which are determined by these tests are listed in Table~\ref{tab:systmeanpt}. Another source of systematic uncertainty due to \acceff is obtained by varying the input \pt and \y shapes to the MC which is used to determine the \acceff (\pt,\y) map. Weight factors have been estimated using normalized \pt and \y distributions from low, high and integrated multiplicity data samples. Data were divided in sub-samples according to their multiplicity. The systematic uncertainty is evaluated by considering the difference observed on the \acceff corrected \meanptjpsi when re-weighted \acceff (\pt,\y) maps are used instead of the default ones. The systematic uncertainty due to \acceff is found to be negligible in low multiplicity intervals, whereas 2$\%$ and 3$\%$ are taken at the highest multiplicity interval for \s $=$ 5.02 and 13 TeV with a conservative approach. The results are obtained with \meanptjpsi as a constant parameter, and with the piece-wise function of \meanptjpsi ($m_{\rm{\mu^{+}\mu^{-}}}$) (see Section~\ref{sec:meanpt}). The difference in values obtained from both the assumptions are taken as systematic uncertainty. The uncertainty on \meanptjpsi as a function of invariant mass is 0.4$\%$ (0.7$\%$) at \s $=$ 5.02 (13) TeV. The systematic uncertainties due to the efficiencies of tracking, trigger, and matching of track to trigger are correlated among the multiplicity classes~\cite{Acharya:2020giw} and the multiplicity integrated values are obtained from Ref.~\cite{Acharya:2017hjh}. The systematic uncertainty due to signal extraction of \meanptjpsi and \acceff variation with event multiplicity are considered as uncorrelated, hence these affect the relative \meanptjpsi measurement. 


\section{Results and discussion}
\label{result_diss}
\begin{figure}[ht!]
\begin{center}
\includegraphics[height=13.cm,width=11.1cm]{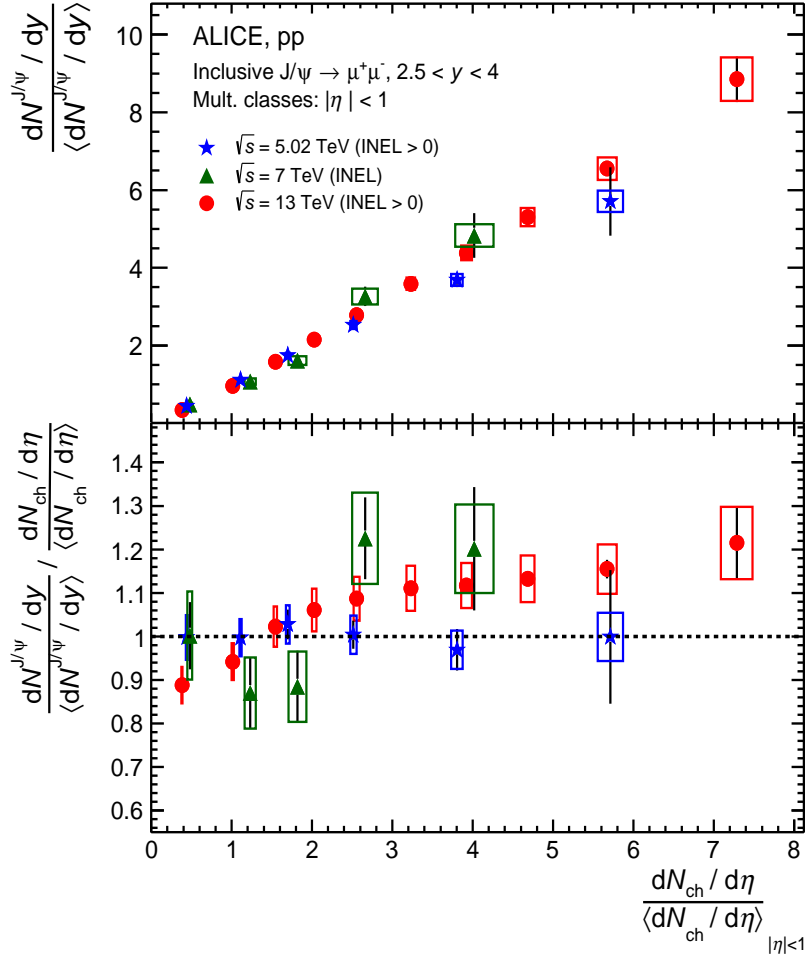}
\end{center}
\caption{Relative J/$\psi$ yield as a function of the relative charged-particle density measured at forward rapidity in \pp collisions at \s $=$ 5.02 (INEL$>$0), 7 (INEL) and 13 (INEL$>$0) TeV. The bottom panel shows the ratio of relative yield of $\rm{J}/\psi$ to relative charged-particle density. The vertical bars and the boxes represent the statistical and the systematic uncertainties, respectively. }
\label{fig:yield}
\end{figure}

The measured relative $\rm{J}/\psi$ yield at forward rapidity (2.5 $<$ \y $<$ 4.0) as a function of \avgdnchdeta at \s $=$ 5.02 and 13 TeV is presented in the upper panel of Fig~\ref{fig:yield}.  At both energies, the relative $J/\psi$ yield shows approximately linear increase with midrapidity relative charged-particle multiplicity. Also in Fig.~\ref{fig:yield}, these results are compared with those from a previous measurement at \s $=$ 7 TeV~\cite{Abelev:2012rz}, in which a close-to-linear trend was observed as well, albeit with significantly higher uncertainties. It is to be noted here that $J/\psi$ yield measurements at \s $=$ 7 TeV are for INEL events and the effect of INEL to INEL $>$ 0 is found to be below 1\%.
The similarity of the results at various collision energies suggests that in a same final state multiplicity domain, the $\rm{J}/\psi$ production at forward rapidity depends to a lesser extent of \s. 
In order to assess better possible deviations from a linear behaviour, the ratio of the relative yield of $\rm{J}/\psi$ to the relative charged-particle multiplicity density is shown as a function of relative charged-particle multiplicity in the bottom panel of Fig.~\ref{fig:yield}. The data points at the three collision energies are compatible with each other and the data points at $\sqrt{s} =$ 5.02 and 7 TeV are also compatible with unity within the uncertainties. The points at $\sqrt{s} =$ 13 TeV are above unity for \avgdnchdeta $\geq$ 1.6. This hints at a 4.9$\sigma$ departure from the linear behaviour for \avgdnchdeta $\geq$ 3 in the case of $\sqrt{s} =$ 13 TeV, although the larger uncertainties at \s $=$ 5.02 and 7 TeV do not allow to exclude a similar behaviour at the lower collision energies.  

\begin{figure}[h!]
\begin{center}
\includegraphics[height=11.cm,width=14.05cm]{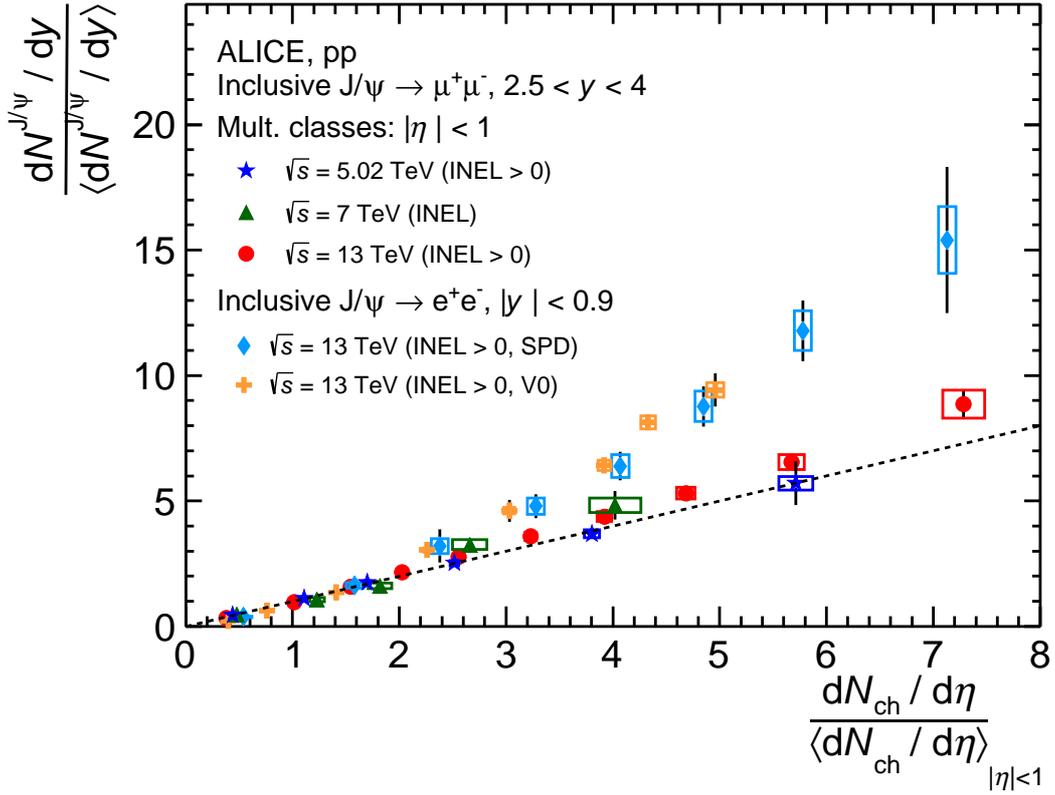}
\end{center}
\caption{ Relative \jpsi yields measured at forward rapidity at \s = 5.02, 7 and 13 TeV compared with similar measurements at midrapidity at \s = 13 TeV,~the latter corresponding to the event selection based on SPD tracklets at midrapidity and on V0 amplitude at forward rapidity.}
\label{fig:compare_allyield_pp}
\end{figure}

In Fig.~\ref{fig:compare_allyield_pp}, the results are also compared with midrapidity measurements at \s  $=$ 13 TeV. The midrapidity relative $\rm{J}/\psi$ yield exhibits faster than linear increase as a function of midrapidity relative charged-particle multiplicity. The results using midrapidity multiplicity selection based on the SPD detector ($|\eta|<1$) and forward-rapidity multiplicity selection based on the V0 detector ($-3.7<\eta<-1.7$ and $2.8<\eta<5.1$) are found to be compatible within the uncertainties. Therefore, the different trends in the multiplicity dependence of the J/$\psi$ production observed at midrapidity and forward rapidity are not due to a possible auto-correlation bias, arising from the multiplicity selection.

\begin{figure}[ht!]
\begin{center}
\includegraphics[scale=0.6]{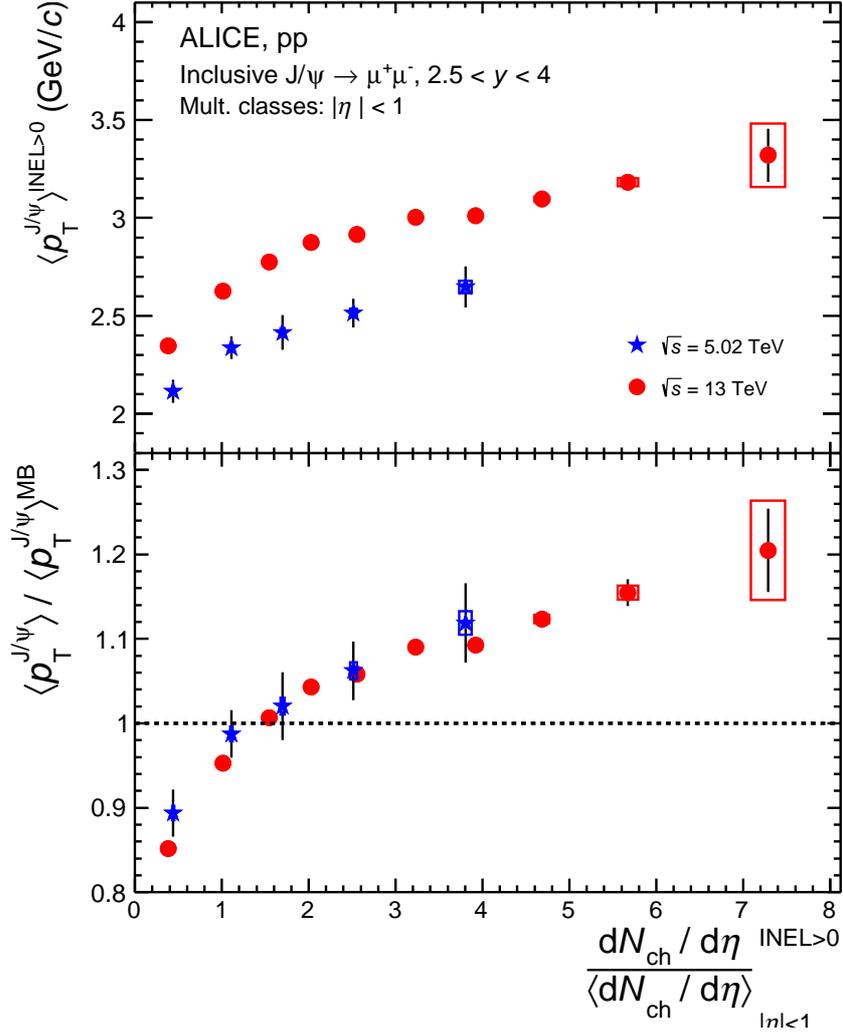}
\end{center}
\caption{ (Top panel) Forward rapidity mean transverse momentum of J$/\psi$ as a function of the relative charged-particle density in \pp collisions at \s $=$ 5.02 and 13 TeV. (Bottom panel) Relative \meanptjpsi shown for the same. The dotted line indicates the deviation of relative \meanptjpsi from one.}
\label{fig:mpT}
\end{figure}

The \meanptjpsi as a function of the relative charged-particle multiplicity measured in \pp collisions at \s $=$ 5.02 and 13 TeV is shown in Fig.~\ref{fig:mpT}. At both the collision energies, an increase of \meanptjpsi with \avgdnchdeta is observed with a possible saturation towards high multiplicity. A similar increase is observed in case of charged-particle \meanpt in \pp collisions~\cite{Abelev:2013bla}. Within the PYTHIA event generator, this increase is due to the Color Reconnection (CR) mechanism, which represents a fusion of hadronizing final-state strings produced in MPI~\cite{Ortiz:2013yxa,ATL-PHYS-PUB-2017-008}. It is worth noting that in the absence of CR, the incoherent superposition of MPI leads to a flat behaviour of \meanpt at high multiplicities. The EPOS event generator~\cite{Drescher:2000ha,Werner:2013tya} is also able to qualitatively reproduce the charged particle \meanpt as a function of charged-particle multiplicity~\cite{Pierog:2013ria}. The generator is based on a combination of Gribov-Regge theory and perturbative QCD, and takes into account multiple parton interactions, non-linear effects (via saturation scales), as well as a 3D+1 viscous hydrodynamical evolution starting from flux tube initial conditions. Within EPOS, the charged-particle \meanpt increases as a function of the multiplicity as a consequence of the collective hadronization of the high-density core produced in the collisions and the increasing relative contribution of the core to the particle production at high multiplicities.
Figure~\ref{fig:mpT} also shows that the absolute \meanptjpsi at $\sqrt{s} =$ 13 TeV is higher than that at \s $=$ 5.02 TeV, as expected from the observed hardening of the corresponding transverse momentum distributions with increasing \s~\cite{Acharya:2017hjh}. However, the relative \meanptjpsi, defined as the ratio of \meanptjpsi in a multiplicity interval to that of minimum bias, is found to be consistent between both collision energies~(Fig.~\ref{fig:mpT}). 

\section{Comparison with models}
\label{modelcomp}
\begin{figure}[ht!]
\subfigure{\includegraphics[scale=0.82]{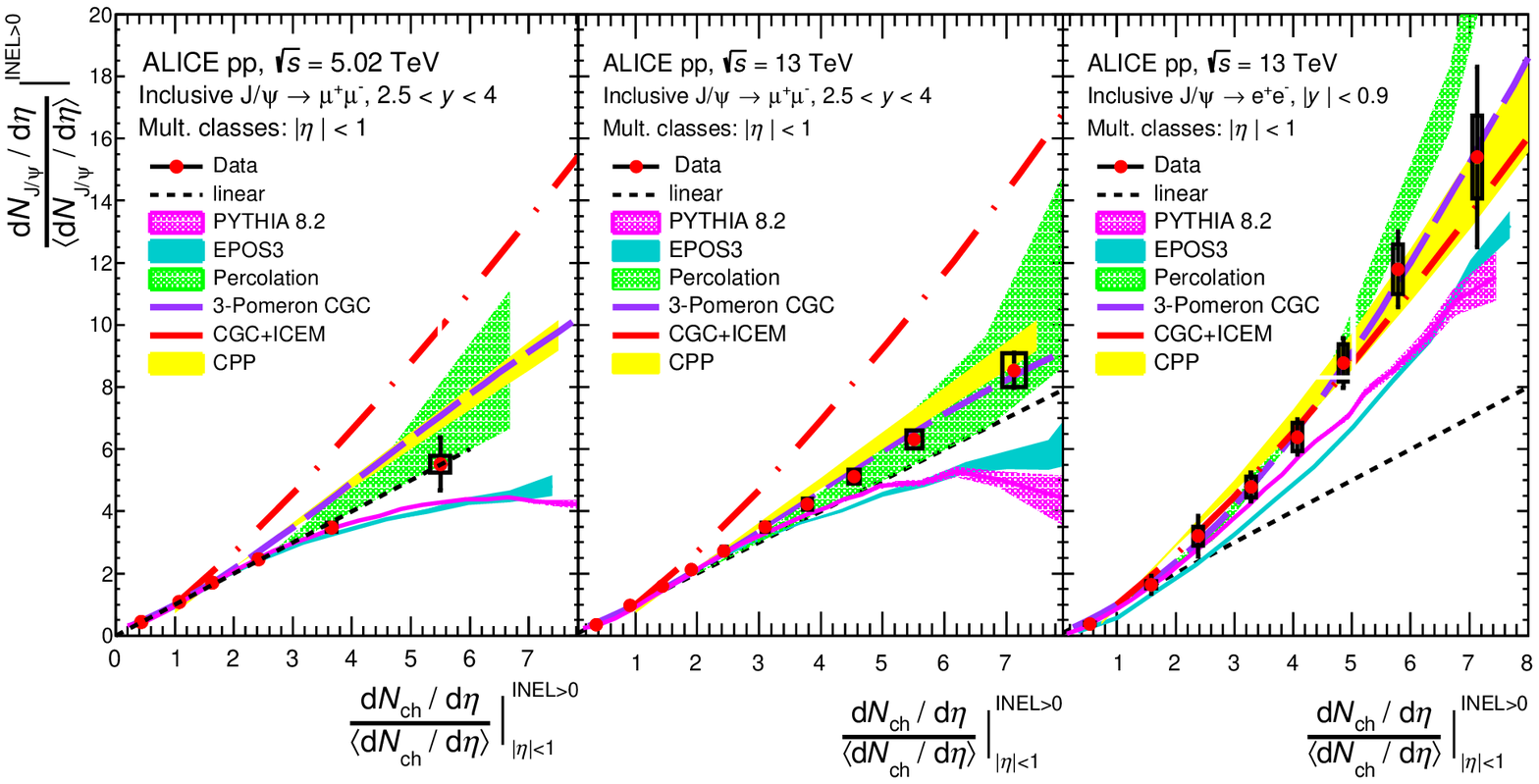}}
\caption{ Comparison of the relative $\rm{J}/\psi$ yield as a function of the relative charged-particle density with model predictions: CPP~\cite{Kopeliovich:2019phc}, CGC with ICEM~\cite{Ma:2018bax}, 3-Pomeron CGC~\cite{Siddikov:2019xvf}, Percolation~\cite{Ferreiro:2012fb}, EPOS3 event generator~\cite{Werner:2013tya}, and PYTHIA 8.2~\cite{Sjostrand:2014zea} at forward rapidity in \pp collisions at \s $=$ 5.02 (left) and 13 TeV (middle). The right hand side figure shows the results from midrapidity compared to the corresponding theoretical model estimations for pp collisions at \s $=$ 13 TeV~\cite{Acharya:2020pit}.}
\label{fig:RelyieldModel}
\end{figure}

Fig.~\ref{fig:RelyieldModel} shows the comparison between the measured relative J$/\psi$ yield as a function of multiplicity and theoretical calculations for $\sqrt{s} =$ 5.02 and 13 TeV at forward rapidity. The Coherent Particle Production (CPP)~\cite{Kopeliovich:2019phc}, CGC with ICEM (improved color evaporation model)~\cite{Ma:2018bax}, 3-Pomeron CGC~\cite{Siddikov:2019xvf}, Percolation~\cite{Ferreiro:2012fb}, EPOS3 event generator~\cite{Werner:2013tya}, and PYTHIA 8.2 (Monash 2013)~\cite{Sjostrand:2014zea} models are represented by the blue shaded region, red dotted line, dashed magenta line, grey shaded region, green shaded region, and red shaded region, respectively.

The EPOS3~\cite{Werner:2013tya} and PYTHIA 8.2~\cite{Sjostrand:2014zea} event generators, already introduced in Section~\ref{result_diss}, predict slightly lower than linear increase of the relative J/$\psi$ yield at forward rapidity as a function of relative multiplicity. Thus, they are able to describe the data at low multiplicity, but significantly underestimate the data at high multiplicity. It is worth noting that similar underestimation of the data is present at midrapidity~\cite{Acharya:2020pit}. 

The CPP model~\cite{Kopeliovich:2019phc,Kopeliovich:2013yfa} relies on the phenomenological parametrization for mean multiplicities of light hadrons and J/$\psi$. The parametrization is derived from p--Pb collisions assuming a linear dependence on binary nucleon--nucleon collisions ($\rm{N}_{coll}$). Within the model, the nuclear suppression is found to depend weakly on energy~\cite{Kopeliovich:2017jpy} and therefore the used parametrization is independent of center-of-mass energy. An excellent agreement is found between the model and the measurements in high-multiplicity events at both collision energies. 

The Color Glass Condensate (CGC) approach with ICEM employs the NRQCD framework in order to describe the J/$\psi$ hadronisation process~\cite{Ma:2018bax}. According to this model, the heavy-flavour production in high-multiplicity events in pp and pA collisions is sensitive to strongly correlated gluons in the colliding protons and nuclei. The dynamics of such configurations is controlled by semihard saturation scale $Q_{s}$ ($x$) in each of the colliding hadrons, where $x$ is the longitudinal momentum fraction carried by a parton in the hadron. $Q_{s}$ increases with decreasing $x$ and increasing nuclear size. The model predicts a faster than linear increase of J/$\psi$ yield with multiplicity in both pp and p--Pb collisions. Although the model describes well the midrapidity J/$\psi$ results~\cite{Acharya:2020pit}, it overestimates the forward rapidity measurements, which can be clearly seen from Fig.~\ref{fig:RelyieldModel}. The 3-gluon fusion model~\cite{Siddikov:2019xvf} assumes that the dominant contribution comes from gluon--gluon fusion and the heavy quarks formed in the process might emit soft gluons in order to form quarkonium states. 
It is found that two-gluon fusion mechanism significantly underestimates the multiplicity dependence of J/$\psi$ production~\cite{Siddikov:2019xvf}, whereas the 3-gluon fusion model reproduces well the measured J/$\psi$ yields at both midrapidity~\cite{Acharya:2020pit} and forward rapidity (Fig.~\ref{fig:RelyieldModel}) in pp collisions at \s = 13 TeV.
 
The percolation model~\cite{Ferreiro:2012fb} considers high-energy hadronic collisions as driven by the exchange of color sources (strings) between the projectile and the target. The strings have finite spatial extent and thus can interact reducing their effective number. One can distinguish between soft and hard strings, depending on their transverse masses, i.e. their quark compositions and their transverse momenta. The string transverse size ($r_{T}$) is determined by its transverse mass ($m_{\rm T}$), since $r_{\rm T} = 1/m_{\rm T}$. The softness of the source maximizes its possibility of interaction, as its transverse size is larger. At high densities, the coherence among the sources (partons or strings) leads to a reduction of their effective number, initially proportional to the number of parton--parton interactions. This reduction mainly affects the soft observables, such as the total multiplicity, while hard production remains unaltered. At low multiplicities, with smaller number of strings the dependence of relative J/$\psi$ yield with relative charged-particle multiplicity is linear, whereas the linear dependence changes to a quadratic dependence for high multiplicity. The number of strings, is smaller at forward rapidity compared to midrapidity and therefore the predicted rise of relative J/$\psi$ yield with relative charged-particle multiplicity at forward rapidity is slower compared to the midrapidity one, which is in agreement with the corresponding measurements. Fig.~\ref{fig:RelyieldModel} shows that the percolation model explains well the data within theoretical uncertainties at both $\sqrt{s} =$ 5.02 and 13 TeV.

\begin{figure}[ht!]
\begin{center}
\includegraphics[scale=0.7]{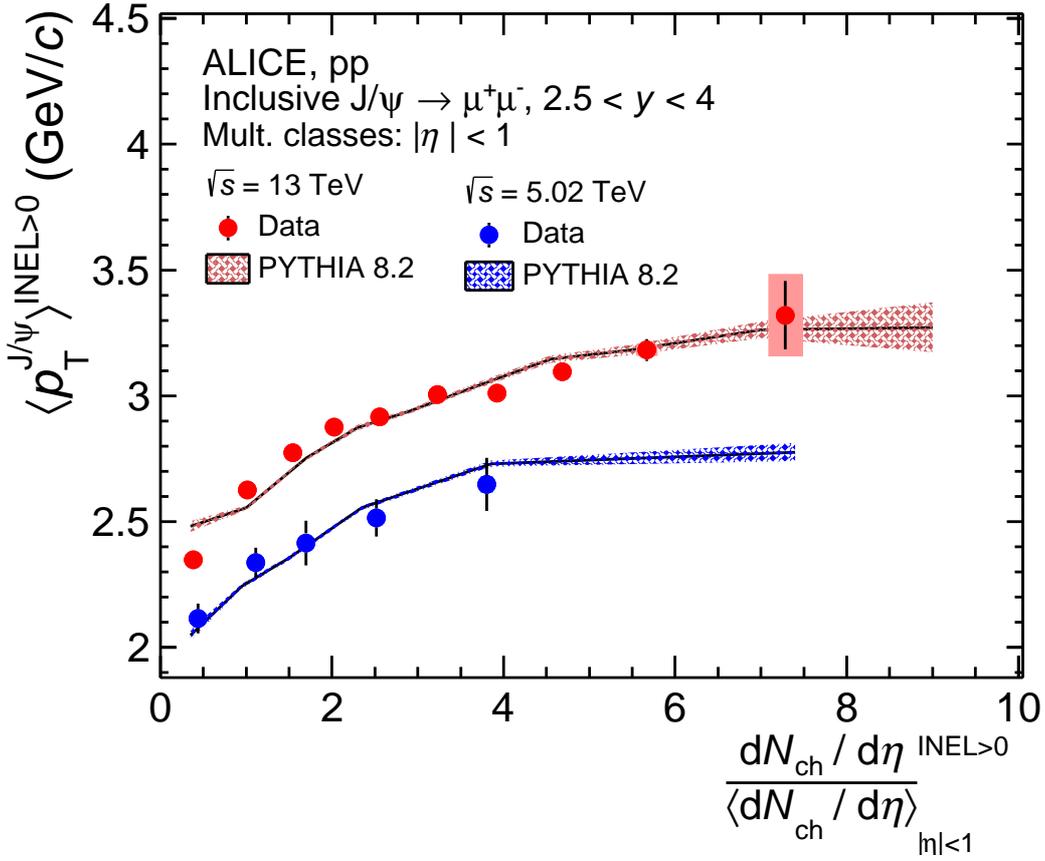}
\end{center}
\caption{Comparison of forward rapidity \meanptjpsi as a function of the relative charged-particle density in \pp collisions at \s $=$ 5.02 and 13 TeV with PYTHIA 8.2~\cite{Sjostrand:2014zea}.}
\label{fig:mpTmodel}
\end{figure}

Figure~\ref{fig:mpTmodel} shows that within uncertainties, PYTHIA8 (Monash 2013) with color reconnection gives a reasonable description of the multiplicity dependence of \meanptjpsi. \meanpt as estimated within the framework of PYTHIA8 is closer to the data in the discussed multiplicity regime, except the lowest multiplicity for \s $=$ 13 TeV data. Both for data and  PYTHIA8, the average \meanpt increases at a constant rate with multiplicity for \avgdnchdeta $>$ 2.5 and \avgdnchdeta $>$ 4.5 for \s $=$ 5.02 and 13 TeV, respectively.

\section{Summary}
\label{summary}

The multiplicity dependence of $\rm{J}/\psi$ production at forward rapidity has been measured as a function of charged-particle multiplicity at midrapidity in $\rm{INEL>0}$ pp collisions at \s $=$ 5.02 and 13 TeV. The present measurement extends the multiplicity reach with respect to the previous measurements at \s $=$ 7 TeV. The results exhibit an approximately linear dependence of the $\rm{J}/\psi$ yield as a function of the event multiplicity, independent of the collision energy. All the discussed models qualitatively describe the observed trend, however the 3-Pomeron CGC model, percolation model and the CPP model quantitatively reproduce the results.

The first moment of the transverse momentum of the \jpsi at forward rapidity as a function of the charged-particle multiplicity is explored for the first time in pp collisions at \s $=$ 5.02 and 13 TeV. 
The \meanptjpsi is found to increase with the event multiplicity with a hint of a saturation towards high multiplicity. The absolute \meanptjpsi at \s $=$ 13 TeV is higher than that at \s $=$ 5.02 TeV, while the relative \meanptjpsi is consistent between both collision energies.


\newenvironment{acknowledgement}{\relax}{\relax}
\begin{acknowledgement}
\section*{Acknowledgements}
The authors would like to acknowledge Elena Gonzalez Ferreiro, Boris Kopeliovich, Klaus Werner, Kazuhiro Watanabe,  Marat Siddikov for useful discussions and for providing theoretical model calculations used for data comparison.

The ALICE Collaboration would like to thank all its engineers and technicians for their invaluable contributions to the construction of the experiment and the CERN accelerator teams for the outstanding performance of the LHC complex.
The ALICE Collaboration gratefully acknowledges the resources and support provided by all Grid centres and the Worldwide LHC Computing Grid (WLCG) collaboration.
The ALICE Collaboration acknowledges the following funding agencies for their support in building and running the ALICE detector:
A. I. Alikhanyan National Science Laboratory (Yerevan Physics Institute) Foundation (ANSL), State Committee of Science and World Federation of Scientists (WFS), Armenia;
Austrian Academy of Sciences, Austrian Science Fund (FWF): [M 2467-N36] and Nationalstiftung f\"{u}r Forschung, Technologie und Entwicklung, Austria;
Ministry of Communications and High Technologies, National Nuclear Research Center, Azerbaijan;
Conselho Nacional de Desenvolvimento Cient\'{\i}fico e Tecnol\'{o}gico (CNPq), Financiadora de Estudos e Projetos (Finep), Funda\c{c}\~{a}o de Amparo \`{a} Pesquisa do Estado de S\~{a}o Paulo (FAPESP) and Universidade Federal do Rio Grande do Sul (UFRGS), Brazil;
Ministry of Education of China (MOEC) , Ministry of Science \& Technology of China (MSTC) and National Natural Science Foundation of China (NSFC), China;
Ministry of Science and Education and Croatian Science Foundation, Croatia;
Centro de Aplicaciones Tecnol\'{o}gicas y Desarrollo Nuclear (CEADEN), Cubaenerg\'{\i}a, Cuba;
Ministry of Education, Youth and Sports of the Czech Republic, Czech Republic;
The Danish Council for Independent Research | Natural Sciences, the VILLUM FONDEN and Danish National Research Foundation (DNRF), Denmark;
Helsinki Institute of Physics (HIP), Finland;
Commissariat \`{a} l'Energie Atomique (CEA) and Institut National de Physique Nucl\'{e}aire et de Physique des Particules (IN2P3) and Centre National de la Recherche Scientifique (CNRS), France;
Bundesministerium f\"{u}r Bildung und Forschung (BMBF) and GSI Helmholtzzentrum f\"{u}r Schwerionenforschung GmbH, Germany;
General Secretariat for Research and Technology, Ministry of Education, Research and Religions, Greece;
National Research, Development and Innovation Office, Hungary;
Department of Atomic Energy Government of India (DAE), Department of Science and Technology, Government of India (DST), University Grants Commission, Government of India (UGC) and Council of Scientific and Industrial Research (CSIR), India;
Indonesian Institute of Science, Indonesia;
Istituto Nazionale di Fisica Nucleare (INFN), Italy;
Japanese Ministry of Education, Culture, Sports, Science and Technology (MEXT) and Japan Society for the Promotion of Science (JSPS) KAKENHI, Japan;
Consejo Nacional de Ciencia (CONACYT) y Tecnolog\'{i}a, through Fondo de Cooperaci\'{o}n Internacional en Ciencia y Tecnolog\'{i}a (FONCICYT) and Direcci\'{o}n General de Asuntos del Personal Academico (DGAPA), Mexico;
Nederlandse Organisatie voor Wetenschappelijk Onderzoek (NWO), Netherlands;
The Research Council of Norway, Norway;
Commission on Science and Technology for Sustainable Development in the South (COMSATS), Pakistan;
Pontificia Universidad Cat\'{o}lica del Per\'{u}, Peru;
Ministry of Education and Science, National Science Centre and WUT ID-UB, Poland;
Korea Institute of Science and Technology Information and National Research Foundation of Korea (NRF), Republic of Korea;
Ministry of Education and Scientific Research, Institute of Atomic Physics, Ministry of Research and Innovation and Institute of Atomic Physics and University Politehnica of Bucharest, Romania;
Joint Institute for Nuclear Research (JINR), Ministry of Education and Science of the Russian Federation, National Research Centre Kurchatov Institute, Russian Science Foundation and Russian Foundation for Basic Research, Russia;
Ministry of Education, Science, Research and Sport of the Slovak Republic, Slovakia;
National Research Foundation of South Africa, South Africa;
Swedish Research Council (VR) and Knut \& Alice Wallenberg Foundation (KAW), Sweden;
European Organization for Nuclear Research, Switzerland;
Suranaree University of Technology (SUT), National Science and Technology Development Agency (NSDTA) and Office of the Higher Education Commission under NRU project of Thailand, Thailand;
Turkish Energy, Nuclear and Mineral Research Agency (TENMAK), Turkey;
National Academy of  Sciences of Ukraine, Ukraine;
Science and Technology Facilities Council (STFC), United Kingdom;
National Science Foundation of the United States of America (NSF) and United States Department of Energy, Office of Nuclear Physics (DOE NP), United States of America.
\end{acknowledgement}

\bibliographystyle{utphys}   
\bibliography{intro_bib,bibliography}

\newpage
\appendix

%
%

\section{The ALICE Collaboration}
\label{app:collab}
\small
\begin{flushleft} 

S.~Acharya$^{\rm 142}$, 
D.~Adamov\'{a}$^{\rm 96}$, 
A.~Adler$^{\rm 74}$, 
J.~Adolfsson$^{\rm 81}$, 
G.~Aglieri Rinella$^{\rm 34}$, 
M.~Agnello$^{\rm 30}$, 
N.~Agrawal$^{\rm 54}$, 
Z.~Ahammed$^{\rm 142}$, 
S.~Ahmad$^{\rm 16}$, 
S.U.~Ahn$^{\rm 76}$
% ...

\section*{Affiliation Notes}

$^{\rm I}$ Deceased\\
$^{\rm II}$ Also at: Italian National Agency for New Technologies, Energy and Sustainable Economic Development (ENEA), Bologna, Italy\\
$^{\rm III}$ Also at: Dipartimento DET del Politecnico di Torino, Turin, Italy\\
$^{\rm IV}$ Also at: M.V. Lomonosov Moscow State University, D.V. Skobeltsyn Institute of Nuclear, Physics, Moscow, Russia\\
$^{\rm V}$ Also at: Department of Applied Physics, Aligarh Muslim University, Aligarh, India\\
$^{\rm VI}$ Also at: Institute of Theoretical Physics, University of Wroclaw, Poland\\
$^{\rm VII}$ Also at: University of Kansas, Lawrence, Kansas, United States\\

\section*{Collaboration Institutes}

$^{1}$ A.I. Alikhanyan National Science Laboratory (Yerevan Physics Institute) Foundation, Yerevan, Armenia\\
$^{2}$ AGH University of Science and Technology, Cracow, Poland\\
$^{3}$ Bogolyubov Institute for Theoretical Physics, National Academy of Sciences of Ukraine, Kiev, Ukraine\\
$^{4}$ Bose Institute, Department of Physics  and Centre for Astroparticle Physics and Space Science (CAPSS), Kolkata, India\\
$^{5}$ Budker Institute for Nuclear Physics, Novosibirsk, Russia\\
$^{6}$ California Polytechnic State University, San Luis Obispo, California, United States\\
$^{7}$ Central China Normal University, Wuhan, China\\
$^{8}$ Centro de Aplicaciones Tecnol\'{o}gicas y Desarrollo Nuclear (CEADEN), Havana, Cuba\\
$^{9}$ Centro de Investigaci\'{o}n y de Estudios Avanzados (CINVESTAV), Mexico City and M\'{e}rida, Mexico\\
$^{10}$ Chicago State University, Chicago, Illinois, United States\\
$^{11}$ China Institute of Atomic Energy, Beijing, China\\
$^{12}$ Chungbuk National University, Cheongju, Republic of Korea\\
$^{13}$ Comenius University Bratislava, Faculty of Mathematics, Physics and Informatics, Bratislava, Slovakia\\
$^{14}$ COMSATS University Islamabad, Islamabad, Pakistan\\
$^{15}$ Creighton University, Omaha, Nebraska, United States\\
$^{16}$ Department of Physics, Aligarh Muslim University, Aligarh, India\\
$^{17}$ Department of Physics, Pusan National University, Pusan, Republic of Korea\\
$^{18}$ Department of Physics, Sejong University, Seoul, Republic of Korea\\
$^{19}$ Department of Physics, University of California, Berkeley, California, United States\\
$^{20}$ Department of Physics, University of Oslo, Oslo, Norway\\
$^{21}$ Department of Physics and Technology, University of Bergen, Bergen, Norway\\
$^{22}$ Dipartimento di Fisica dell'Universit\`{a} and Sezione INFN, Cagliari, Italy\\
$^{23}$ Dipartimento di Fisica dell'Universit\`{a} and Sezione INFN, Trieste, Italy\\
$^{24}$ Dipartimento di Fisica dell'Universit\`{a} and Sezione INFN, Turin, Italy\\
$^{25}$ Dipartimento di Fisica e Astronomia dell'Universit\`{a} and Sezione INFN, Bologna, Italy\\
$^{26}$ Dipartimento di Fisica e Astronomia dell'Universit\`{a} and Sezione INFN, Catania, Italy\\
$^{27}$ Dipartimento di Fisica e Astronomia dell'Universit\`{a} and Sezione INFN, Padova, Italy\\
$^{28}$ Dipartimento di Fisica e Nucleare e Teorica, Universit\`{a} di Pavia, Pavia, Italy\\
$^{29}$ Dipartimento di Fisica `E.R.~Caianiello' dell'Universit\`{a} and Gruppo Collegato INFN, Salerno, Italy\\
$^{30}$ Dipartimento DISAT del Politecnico and Sezione INFN, Turin, Italy\\
$^{31}$ Dipartimento di Scienze e Innovazione Tecnologica dell'Universit\`{a} del Piemonte Orientale and INFN Sezione di Torino, Alessandria, Italy\\
$^{32}$ Dipartimento di Scienze MIFT, Universit\`{a} di Messina, Messina, Italy\\
$^{33}$ Dipartimento Interateneo di Fisica `M.~Merlin' and Sezione INFN, Bari, Italy\\
$^{34}$ European Organization for Nuclear Research (CERN), Geneva, Switzerland\\
$^{35}$ Faculty of Electrical Engineering, Mechanical Engineering and Naval Architecture, University of Split, Split, Croatia\\
$^{36}$ Faculty of Engineering and Science, Western Norway University of Applied Sciences, Bergen, Norway\\
$^{37}$ Faculty of Nuclear Sciences and Physical Engineering, Czech Technical University in Prague, Prague, Czech Republic\\
$^{38}$ Faculty of Science, P.J.~\v{S}af\'{a}rik University, Ko\v{s}ice, Slovakia\\
$^{39}$ Frankfurt Institute for Advanced Studies, Johann Wolfgang Goethe-Universit\"{a}t Frankfurt, Frankfurt, Germany\\
$^{40}$ Fudan University, Shanghai, China\\
$^{41}$ Gangneung-Wonju National University, Gangneung, Republic of Korea\\
$^{42}$ Gauhati University, Department of Physics, Guwahati, India\\
$^{43}$ Helmholtz-Institut f\"{u}r Strahlen- und Kernphysik, Rheinische Friedrich-Wilhelms-Universit\"{a}t Bonn, Bonn, Germany\\
$^{44}$ Helsinki Institute of Physics (HIP), Helsinki, Finland\\
$^{45}$ High Energy Physics Group,  Universidad Aut\'{o}noma de Puebla, Puebla, Mexico\\
$^{46}$ Hiroshima University, Hiroshima, Japan\\
$^{47}$ Hochschule Worms, Zentrum  f\"{u}r Technologietransfer und Telekommunikation (ZTT), Worms, Germany\\
$^{48}$ Horia Hulubei National Institute of Physics and Nuclear Engineering, Bucharest, Romania\\
$^{49}$ Indian Institute of Technology Bombay (IIT), Mumbai, India\\
$^{50}$ Indian Institute of Technology Indore, Indore, India\\
$^{51}$ Indonesian Institute of Sciences, Jakarta, Indonesia\\
$^{52}$ INFN, Laboratori Nazionali di Frascati, Frascati, Italy\\
$^{53}$ INFN, Sezione di Bari, Bari, Italy\\
$^{54}$ INFN, Sezione di Bologna, Bologna, Italy\\
$^{55}$ INFN, Sezione di Cagliari, Cagliari, Italy\\
$^{56}$ INFN, Sezione di Catania, Catania, Italy\\
$^{57}$ INFN, Sezione di Padova, Padova, Italy\\
$^{58}$ INFN, Sezione di Pavia, Pavia, Italy\\
$^{59}$ INFN, Sezione di Torino, Turin, Italy\\
$^{60}$ INFN, Sezione di Trieste, Trieste, Italy\\
$^{61}$ Inha University, Incheon, Republic of Korea\\
$^{62}$ Institute for Gravitational and Subatomic Physics (GRASP), Utrecht University/Nikhef, Utrecht, Netherlands\\
$^{63}$ Institute for Nuclear Research, Academy of Sciences, Moscow, Russia\\
$^{64}$ Institute of Experimental Physics, Slovak Academy of Sciences, Ko\v{s}ice, Slovakia\\
$^{65}$ Institute of Physics, Homi Bhabha National Institute, Bhubaneswar, India\\
$^{66}$ Institute of Physics of the Czech Academy of Sciences, Prague, Czech Republic\\
$^{67}$ Institute of Space Science (ISS), Bucharest, Romania\\
$^{68}$ Institut f\"{u}r Kernphysik, Johann Wolfgang Goethe-Universit\"{a}t Frankfurt, Frankfurt, Germany\\
$^{69}$ Instituto de Ciencias Nucleares, Universidad Nacional Aut\'{o}noma de M\'{e}xico, Mexico City, Mexico\\
$^{70}$ Instituto de F\'{i}sica, Universidade Federal do Rio Grande do Sul (UFRGS), Porto Alegre, Brazil\\
$^{71}$ Instituto de F\'{\i}sica, Universidad Nacional Aut\'{o}noma de M\'{e}xico, Mexico City, Mexico\\
$^{72}$ iThemba LABS, National Research Foundation, Somerset West, South Africa\\
$^{73}$ Jeonbuk National University, Jeonju, Republic of Korea\\
$^{74}$ Johann-Wolfgang-Goethe Universit\"{a}t Frankfurt Institut f\"{u}r Informatik, Fachbereich Informatik und Mathematik, Frankfurt, Germany\\
$^{75}$ Joint Institute for Nuclear Research (JINR), Dubna, Russia\\
$^{76}$ Korea Institute of Science and Technology Information, Daejeon, Republic of Korea\\
$^{77}$ KTO Karatay University, Konya, Turkey\\
$^{78}$ Laboratoire de Physique des 2 Infinis, Ir\`{e}ne Joliot-Curie, Orsay, France\\
$^{79}$ Laboratoire de Physique Subatomique et de Cosmologie, Universit\'{e} Grenoble-Alpes, CNRS-IN2P3, Grenoble, France\\
$^{80}$ Lawrence Berkeley National Laboratory, Berkeley, California, United States\\
$^{81}$ Lund University Department of Physics, Division of Particle Physics, Lund, Sweden\\
$^{82}$ Moscow Institute for Physics and Technology, Moscow, Russia\\
$^{83}$ Nagasaki Institute of Applied Science, Nagasaki, Japan\\
$^{84}$ Nara Women{'}s University (NWU), Nara, Japan\\
$^{85}$ National and Kapodistrian University of Athens, School of Science, Department of Physics , Athens, Greece\\
$^{86}$ National Centre for Nuclear Research, Warsaw, Poland\\
$^{87}$ National Institute of Science Education and Research, Homi Bhabha National Institute, Jatni, India\\
$^{88}$ National Nuclear Research Center, Baku, Azerbaijan\\
$^{89}$ National Research Centre Kurchatov Institute, Moscow, Russia\\
$^{90}$ Niels Bohr Institute, University of Copenhagen, Copenhagen, Denmark\\
$^{91}$ Nikhef, National institute for subatomic physics, Amsterdam, Netherlands\\
$^{92}$ NRC Kurchatov Institute IHEP, Protvino, Russia\\
$^{93}$ NRC \guillemotleft Kurchatov\guillemotright  Institute - ITEP, Moscow, Russia\\
$^{94}$ NRNU Moscow Engineering Physics Institute, Moscow, Russia\\
$^{95}$ Nuclear Physics Group, STFC Daresbury Laboratory, Daresbury, United Kingdom\\
$^{96}$ Nuclear Physics Institute of the Czech Academy of Sciences, \v{R}e\v{z} u Prahy, Czech Republic\\
$^{97}$ Oak Ridge National Laboratory, Oak Ridge, Tennessee, United States\\
$^{98}$ Ohio State University, Columbus, Ohio, United States\\
$^{99}$ Petersburg Nuclear Physics Institute, Gatchina, Russia\\
$^{100}$ Physics department, Faculty of science, University of Zagreb, Zagreb, Croatia\\
$^{101}$ Physics Department, Panjab University, Chandigarh, India\\
$^{102}$ Physics Department, University of Jammu, Jammu, India\\
$^{103}$ Physics Department, University of Rajasthan, Jaipur, India\\
$^{104}$ Physikalisches Institut, Eberhard-Karls-Universit\"{a}t T\"{u}bingen, T\"{u}bingen, Germany\\
$^{105}$ Physikalisches Institut, Ruprecht-Karls-Universit\"{a}t Heidelberg, Heidelberg, Germany\\
$^{106}$ Physik Department, Technische Universit\"{a}t M\"{u}nchen, Munich, Germany\\
$^{107}$ Politecnico di Bari and Sezione INFN, Bari, Italy\\
$^{108}$ Research Division and ExtreMe Matter Institute EMMI, GSI Helmholtzzentrum f\"ur Schwerionenforschung GmbH, Darmstadt, Germany\\
$^{109}$ Russian Federal Nuclear Center (VNIIEF), Sarov, Russia\\
$^{110}$ Saha Institute of Nuclear Physics, Homi Bhabha National Institute, Kolkata, India\\
$^{111}$ School of Physics and Astronomy, University of Birmingham, Birmingham, United Kingdom\\
$^{112}$ Secci\'{o}n F\'{\i}sica, Departamento de Ciencias, Pontificia Universidad Cat\'{o}lica del Per\'{u}, Lima, Peru\\
$^{113}$ St. Petersburg State University, St. Petersburg, Russia\\
$^{114}$ Stefan Meyer Institut f\"{u}r Subatomare Physik (SMI), Vienna, Austria\\
$^{115}$ SUBATECH, IMT Atlantique, Universit\'{e} de Nantes, CNRS-IN2P3, Nantes, France\\
$^{116}$ Suranaree University of Technology, Nakhon Ratchasima, Thailand\\
$^{117}$ Technical University of Ko\v{s}ice, Ko\v{s}ice, Slovakia\\
$^{118}$ The Henryk Niewodniczanski Institute of Nuclear Physics, Polish Academy of Sciences, Cracow, Poland\\
$^{119}$ The University of Texas at Austin, Austin, Texas, United States\\
$^{120}$ Universidad Aut\'{o}noma de Sinaloa, Culiac\'{a}n, Mexico\\
$^{121}$ Universidade de S\~{a}o Paulo (USP), S\~{a}o Paulo, Brazil\\
$^{122}$ Universidade Estadual de Campinas (UNICAMP), Campinas, Brazil\\
$^{123}$ Universidade Federal do ABC, Santo Andre, Brazil\\
$^{124}$ University of Cape Town, Cape Town, South Africa\\
$^{125}$ University of Houston, Houston, Texas, United States\\
$^{126}$ University of Jyv\"{a}skyl\"{a}, Jyv\"{a}skyl\"{a}, Finland\\
$^{127}$ University of Kansas, Lawrence, Kansas, United States\\
$^{128}$ University of Liverpool, Liverpool, United Kingdom\\
$^{129}$ University of Science and Technology of China, Hefei, China\\
$^{130}$ University of South-Eastern Norway, Tonsberg, Norway\\
$^{131}$ University of Tennessee, Knoxville, Tennessee, United States\\
$^{132}$ University of the Witwatersrand, Johannesburg, South Africa\\
$^{133}$ University of Tokyo, Tokyo, Japan\\
$^{134}$ University of Tsukuba, Tsukuba, Japan\\
$^{135}$ University Politehnica of Bucharest, Bucharest, Romania\\
$^{136}$ Universit\'{e} Clermont Auvergne, CNRS/IN2P3, LPC, Clermont-Ferrand, France\\
$^{137}$ Universit\'{e} de Lyon, CNRS/IN2P3, Institut de Physique des 2 Infinis de Lyon, Lyon, France\\
$^{138}$ Universit\'{e} de Strasbourg, CNRS, IPHC UMR 7178, F-67000 Strasbourg, France, Strasbourg, France\\
$^{139}$ Universit\'{e} Paris-Saclay Centre d'Etudes de Saclay (CEA), IRFU, D\'{e}partment de Physique Nucl\'{e}aire (DPhN), Saclay, France\\
$^{140}$ Universit\`{a} degli Studi di Foggia, Foggia, Italy\\
$^{141}$ Universit\`{a} di Brescia, Brescia, Italy\\
$^{142}$ Variable Energy Cyclotron Centre, Homi Bhabha National Institute, Kolkata, India\\
$^{143}$ Warsaw University of Technology, Warsaw, Poland\\
$^{144}$ Wayne State University, Detroit, Michigan, United States\\
$^{145}$ Westf\"{a}lische Wilhelms-Universit\"{a}t M\"{u}nster, Institut f\"{u}r Kernphysik, M\"{u}nster, Germany\\
$^{146}$ Wigner Research Centre for Physics, Budapest, Hungary\\
$^{147}$ Yale University, New Haven, Connecticut, United States\\
$^{148}$ Yonsei University, Seoul, Republic of Korea\\

\end{flushleft} 
  
\end{document}